\documentclass[article,twocolumn,aps,superscriptaddress,longbibliography]{revtex4-1}

\usepackage{graphicx}
\usepackage{dcolumn}
\usepackage{bm}
\usepackage{epstopdf}
\usepackage{float}
\usepackage[colorlinks=true, linkcolor=blue, citecolor=cyan, urlcolor=purple]{hyperref}
\usepackage{amsmath}
\usepackage{xcolor}
\usepackage{comment}
\usepackage[normalem]{ulem}
\graphicspath{{./images/}}

\begin{document}
\title{ Polaronic dressing of bound states}

\author{L. A. Pe\~na Ardila}
\affiliation{Dipartimento di Fisica, Universita di Trieste, Strada Costiera 11, I-34151 Trieste, Italy}
\author{A. Camacho-Guardian}
\affiliation{Instituto de Fisica, Universidad Nacional Autonoma de Mexico, Ciudad de Mexico CP 04510, Mexico}
\email{acamacho@fisica.unam.mx}

\date{\today}
\begin{abstract} Polarons have emerged as a powerful concept across many-fields in physics to study an impurity coupled to a quantum bath. The interplay between impurity physics and the formation of composite objects remains a relevant problem to understand how few- and many-body states are robust towards complex environments and polaron physics. In most cases, impurities are point-like objects. The question we address here is how quasiparticle properties are affected when impurities possess an internal structure. The simplest yet fundamental structure for the impurity is a dimer state. Here, we investigate the polaronic properties of a dimer dressed by the elementary excitations of a bosonic bath.  We solve the two-body impurity-impurity problem to determine the position and broadening of the bound state and consider the polaron dressing using a field-theory approach. We demonstrate the emergence of different dressed dimer regimes, where polaron dressing drives a dimer from a well-defined to an ill-defined bound state. 
\end{abstract}

\maketitle

{\it Introduction.-} The polaron remains a powerful tool to understand the behavior of isolated impurities in complex many-body systems~\cite {landau1948effective}. The versatility of this concept has allowed the polaron to be exploited in several fields in physics spanning from atomic physics~\cite{jorgensen2016observation,hu2016bose}, solid-state~\cite{pekar1969theory,Eagles1963,devreese2009frohlich}, quantum optics~\cite{Grusdt2014,Nielsen2020,Camacho2020}, chemistry~\cite{srimath2020exciton,franchini2021polarons,tao2021momentarily,bourelle2022optical} to nuclear physics~\cite{Kobyakov2016,tajima2024polaronic}. The arrival of ultracold quantum gases provided both a platform to quantum simulate complex many-body systems and to explore inaccessible regimes in solid-state setups, ruled by strong correlations~\cite{jorgensen2016observation,hu2016bose,ardila2019analyzing,yan2020bose,skou2021non,Skou2022,etrych2024universal, rath2013field,Ardila2015impurity,Shchadilova2016,grusdt2018strong,Christensen2015,christianen2023phase,levinsen2021quantum}, thermal fluctuations effects~\cite{Levinsen2017,guenther2018bose,field2020fate,Drescher2021,hryhorchak2023trapped,isaule2024functional}, dipolar and charged polarons~\cite{Ardila2018,Ardila2022N,Astrakharchik2021,Christensen2021,Ding2022,Astrakharchik2022,olivas2024,volosniev2023non}, Efimov physics~\cite{levinsen2015impurity,Sun2017,Sun2017b,christianen2022chemistry,christianen2022bose}, universality~\cite{Yoshida2018,Massignan2021,Yegovtsev2022,Skou2022,atoms9020022,cayla2023observation}, lattice polarons~\cite{colussi2023lattice,ding2023polarons,isaule2024bound,amelio2024polaron,Santiago-Garcia_2023,santiago2024lattice,Dima2024,vashisht2024chiral,amelio2024polaron,Dima2021,Christ2024}, angulons~\cite{Schmidt2015,Schmidt2016}, among others~\cite{nielsen2019critical,Guenther2021,bighin2022impurity,camacho2023polaritons,Mostaan2023}. 

In contrast to the remarkable theoretical and experimental progress in the single-impurity regime, much less is known about systems with two or more impurities. Considerable attention has been given to the theoretical study of mediated interactions~\cite{paredes2024perspective}, which challenge previously well-established formalisms from condensed matter physics~\cite{Bardeen1967,Yu2012}. This has led to the prediction of new forms of mediated interactions~\cite{Camacho2018b,Fujii2022,Drescher2023} and strongly interacting bipolarons arising from the exchange of collective modes~\cite{Naidon2018,Camacho2018a,Will2021,Peto2022,Drescher2023}. Despite experimental efforts, clear evidence of mediated interactions and bipolarons has yet to be observed in the context of the Bose polaron, with only the first experiments signatures for Fermi polarons~\cite{baroni2024mediated}.

Bound states between two impurities are not limited to those formed via the exchange of sound modes in the medium; they may also arise from direct interactions between the impurities. A prime example is the exciton, a bound state between an electron and a hole in a semiconductor, which forms due to direct Coulomb attraction. Electrons and holes can individually couple to lattice phonons, forming electron-polaron and hole-polaron quasiparticles, respectively. However, when combined, the resulting exciton may exhibit properties significantly different from those of its bare constituents (electrons and holes). In this context, understanding the effects of polaron formation on exciton properties has drawn considerable attention, particularly in the framework of Fröhlich polarons~\cite{emin2013polarons}.

\begin{figure}[h]
\centering
\includegraphics[width=\columnwidth]{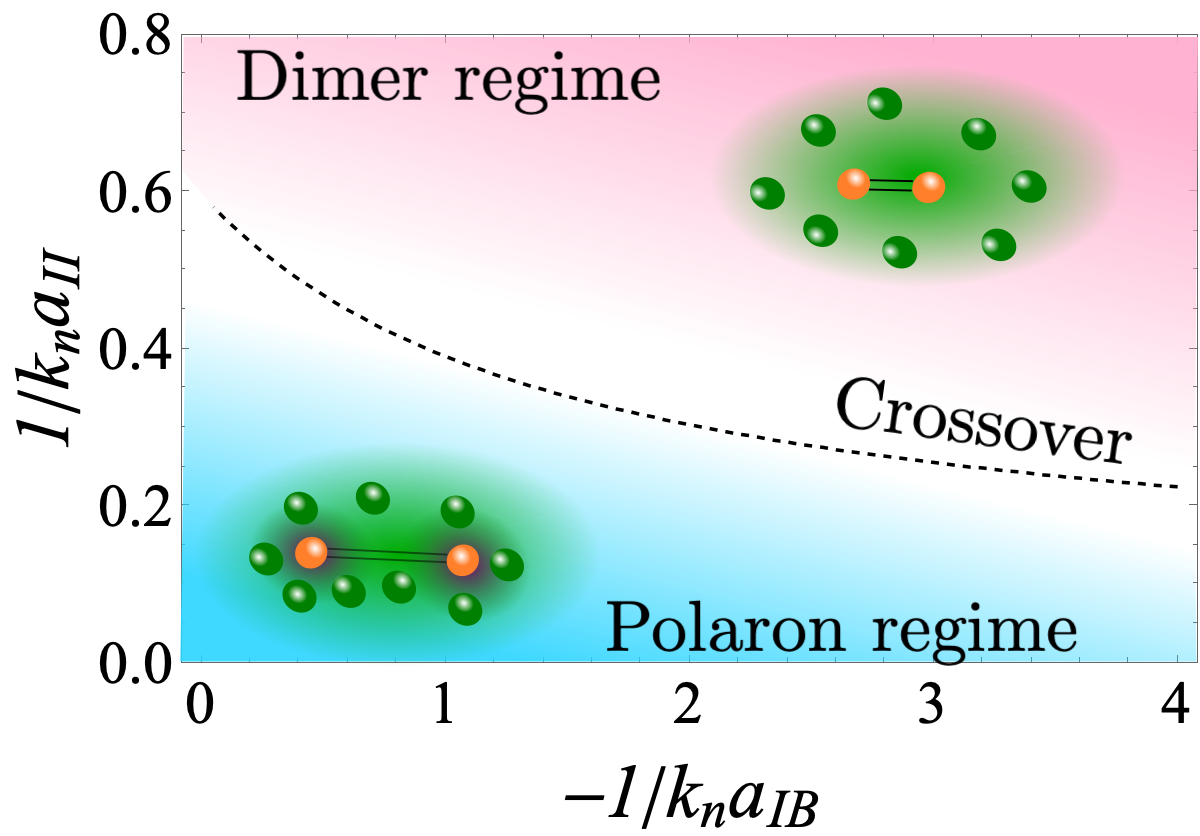}
\caption{Phase diagram of a dressed dimer in a BEC. The interplay between two-body and polaron physics leads to two different regimes: a polaron governed phase and a dimer regime. The phases are  separated by a smooth crossover with no sharp transition.  The dashed line corresponds to $\epsilon_I^{(0)}=\omega_{\mathbf k=0}^P$ when the bare dimer energy matches the polaron energy.}
\label{Fig0}
\end{figure}

To the best of our knowledge, the effects of polaron dressing on existing bound states remain largely unexplored in the strongly interacting regime. Intrigued by this question, in this Article we study the features of a molecular arising from the binding of two atoms strongly coupled to a Bose-Einstein condensate (BEC). We address the two-body problem by solving the Bethe-Salpeter for the interaction of two impurity atoms. The impurity atoms strongly coupled to the BEC forming a polaron treated at the level of the non-self consistent T-matrix approximation. We reveal a phase diagram unveiling an unexplored competition between two-body and polaron physics.

{\it Polaronic dressing of the two-body bound state.-}
 We consider a weakly interacting BEC of $b$ atoms with an equilibrium density $n_0$ at zero temperature $T=0$ and impurities coupled to the BEC. The Hamiltonian of the system, in atomic units, namely $\hbar=1,$ is given by
\begin{gather}
\hat H=\sum_{\mathbf k}(\varepsilon^{(b)}_{\mathbf k}\hat b^\dagger_{\mathbf k}\hat b_{\mathbf k}+\varepsilon^{(c)}_{\mathbf k}\hat c^\dagger_{\mathbf k}\hat c_{\mathbf k})
+g_{IB}\sum_{\mathbf k,\mathbf k',\mathbf q}\hat b^\dagger_{\mathbf k+\mathbf q}\hat c^\dagger_{\mathbf k'-\mathbf q}c_{\mathbf k'}\hat b_{\mathbf k},\\ \nonumber
+\frac{g_{II}}{2}\sum_{\mathbf k,\mathbf k',\mathbf q}\hat c^\dagger_{\mathbf k+\mathbf q}\hat c^\dagger_{\mathbf k'-\mathbf q}c_{\mathbf k'}\hat c_{\mathbf k}.
\end{gather}
here $\hat b^\dagger_{\mathbf k}$ and $\hat c^\dagger_{\mathbf k}$  create a boson/impurity with momentum $\mathbf k$ and energy $\varepsilon^{(b/c)}_{\mathbf k}=\frac{k^2}{2m_{b/c}}.$ The impurity-boson interaction is assumed to be short-ranged and well characterized by coupling strength $g_{IB}=\frac{2\pi a_{IB}}{m_r}$, depending on  the scattering length $a_{IB}$ and the reduced mass $m_{r}^{-1}=m_{b}^{-1}+m_{c}^{-1}$.  The impurity-impurity interaction is also considered to be a contact interaction with $g_{II}=\frac{4\pi a_{II}}{m_c}$. Since we consider a s-wave interaction between the impurities we assume a bosonic statistic. We take the volume of the system $V=1 .$

The direct interaction between impurities can be studied via the Bethe-Salpeter equation (BSE), which gives an exact solution for the two-body problem \cite{Fetter1971}. In the absence of impurity-boson interaction, the impurity-impurity interaction supports a two-body bound state for $a_{II}>0$ with a binding energy of   $\varepsilon_I^{(0)}=-1/m_ca_{II}^2$. The BSE is given by 
\begin{gather}
\Gamma^{-1}(\mathbf Q,\omega)=\frac{m_c}{4\pi a_{II}}-\int \frac{d^3q}{(2\pi)^3}\Pi_{\mathbf q}(\mathbf Q,\omega),
\end{gather}
with the impurity-impurity pair propagator given by
\begin{gather}
\Pi_{\mathbf q}(\mathbf Q,\omega)=\left[\int \frac{d\omega'}{2\pi} S(\mathbf q,\omega')G_{cc}(\mathbf Q-\mathbf q,\omega-\omega')\right]+\frac{1}{2\varepsilon^{(c)}_{\mathbf q}}.
\end{gather}
 Here, the spectral function is defined as $S(\mathbf k,\omega)=-2\text{Im}\mathcal G_{cc}(\mathbf k,\omega).$ In this formalism, the poles of the BSE give the two-body bound states. 

It is important to emphasize that while the BSE provides, in principle, the exact solution for impurity-impurity scattering, it depends on the impurity's Green's function, which can only be obtained approximately in the strong-coupling regime between the impurity and the bath. The impurity Green's function follows Dyson's equation
\begin{gather}
G^{-1}_{cc}(\mathbf k,\omega)=\omega-\frac{k^2}{2m_c}-\Sigma_{cc}(\mathbf k,\omega).
\end{gather}
The problem of the strongly interacting Bose polaron remains an open question. In this article, we focus on understanding the effects of polaron dressing on the formation of two-body bound states. To this end, we employ a well-established theoretical formalism that accurately incorporates Feshbach physics~\cite{rath2013field}, namely the so-called non-self-consistent T-matrix (NSCT) approximation. 

 The analytical expression for the self-energy~\cite{rath2013field}
\begin{gather}
\Sigma_{cc}(\mathbf k,\omega)=n_0\frac{2\pi a_{IB}}{m_r}\frac{1}{1+\frac{i2\pi a_{IB}m_r^{3/2}}{m_r\sqrt{2}\pi}\sqrt{\omega-\frac{k^2}{2M}}},
\end{gather}
Here, we take a vanishing boson-boson interaction $a_{BB}=0,$ for a non-zero boson-boson interaction the qualitative and quantitative quasiparticle properties of the polaron remain very similar in the NSCT approximation~\cite{rath2013field}, and thus, our analysis holds for $a_{BB}>0.$ Here, the total mass is denoted by $M=m_b+m_c$. Under this theoretical approach, the polaron properties can be calculated as following:  the quasiparticle energy $\omega_{\mathbf k}^P$ and residue $Z_{\mathbf k}^P$ are defined in terms of the poles $\text{Re}[\mathcal G^{-1}_{cc}(\mathbf k,\omega_{\mathbf k}^P)]=0$ and $Z_{\mathbf k}^P=[\partial_\omega \text{Re}[\mathcal G^{-1}_{cc}(\mathbf k,\omega)]]_{\omega=\omega_{\mathbf k}^P}^{-1}$ respectively. Further details of the single polaron can be found in a recent review~\cite{Grusdt2024}.

In the absence of polaron dressing, the BSE acquires an analytical expression
\begin{gather}
\label{BS}
\Gamma^{(0)}(\mathbf Q,\omega)=\frac{1}{\frac{m_c}{4\pi a_{II}}+i\frac{m_c^{3/2}}{4\pi}\sqrt{\omega-\frac{Q^2}{4m_c}}}\approx 8\pi\frac{\sqrt{\varepsilon_I^{(0)}}}{ m_c^{3/2}}\frac{1}{\omega-\varepsilon_I^{(0)}},
\end{gather}
where the last expression only holds around the pole of the BSE,  which of course, lies at the energy of the bare dimer $\varepsilon_I^{(0)}=-1/m_ca_{II}^2$. Here, we will fix $m_c=m_b=m.$

Intuitively, we could expect that if the quasiparticle is a good approximation then $G_{cc}(\mathbf k,\omega)\approx Z_{\mathbf k}^{(P)}/(\omega-\omega_{\mathbf k}^P).$ In this case, we expect the bound state to shift to $\varepsilon_I=2\omega_{k=0}-\frac{1}{m_Pa_{\text{eff}}^2}$ where the impurity-impurity scattering length is renormalized by the quasiparticle properties $a_{\text{eff}}=Z_P^2a_{II},$ and $m_c$ by the effective mass of the polaron $m_P.$ In this approach, the impurity-impurity interaction is normalized by the quasiparticle residue, that is, only the coherent part  interact, and thus the effective interaction is suppressed by a factor of $Z_P^2.$ 

{The BSE has been used to study bipolarons arising from the exchange of sound modes~\cite{Camacho2018a}. Due to the inherent complexity of the mediated interaction, which can be non-local and retarded, the BSE has so far been solved only within the quasiparticle approximation for the impurity's Green’s function~\cite{Camacho2018a}. In this work, we go beyond this approximation by retaining the full Green’s function to solve the BSE, allowing us not only to determine the binding energy but also to address the full scattering problem.

}
To understand the polaron effects on the scattering properties we introduce the spectral function of the scattering matrix defined as $A(\mathbf Q,\omega)=-2\text{Im}\Gamma(\mathbf Q,\omega)$ and take first $\mathbf Q=0,$ and solve the BSE  for a fixed value of $a_{II}$ and for varying impurity-boson interactions $a_{IB}.$ \\ 

\begin{figure}[h]
\centering
\includegraphics[width=\columnwidth]{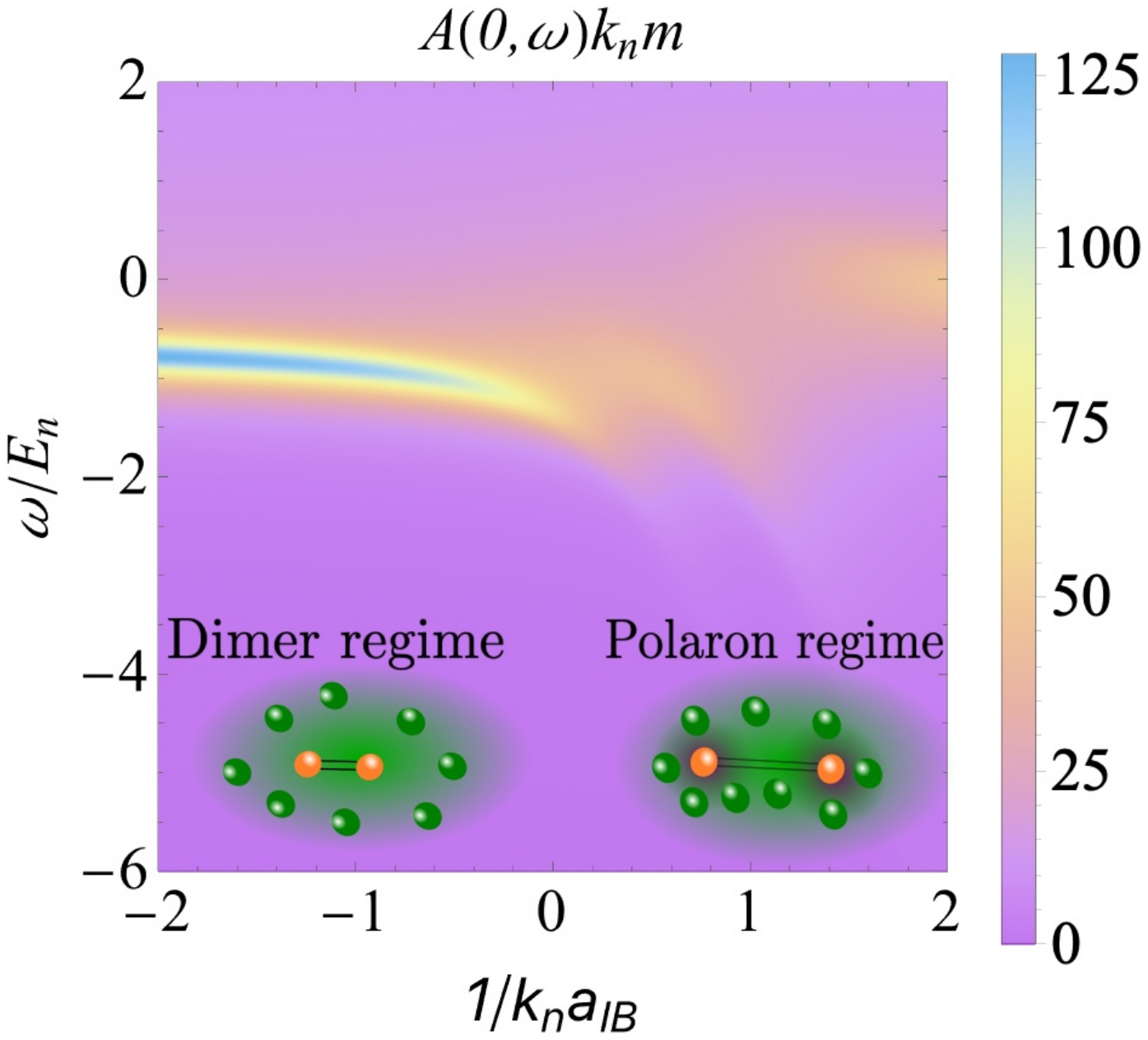}
\caption{The spectral function of the impurity-impurity scattering matrix is plotted as a function of $1/k_n a_{IB}$. The system undergoes a transition from supporting a well-defined dressed dimer in the regime of weak attractive impurity-boson interactions, which broadens with increasing polaron dressing. For strong impurity-boson repulsive interactions, no two-body bound state can be discerned. In this regime, we observe that the spectral function becomes broad and no sharp peak can be distinguished, unveiling an intriguing interplay between a dressed dimer impurity-impurity state and the underlying impurity-boson bound state.  Here we consider $1/k_n a_{II}=0.5$.}
\label{Fig1}
\end{figure}

We first study the case of $1/k_n a_{II}=0.5$, which corresponds to a binding energy of $\varepsilon_I^{(0)}/E_n = -0.5$.  The spectral function of the impurity-impurity scattering matrix is shown in Fig.~\ref{Fig1} as a function of $1/k_na_{IB}$, which illustrates that the \textit{dressed dimer} evolves from a well- to an ill-defined state. For weak boson-impurity interactions the spectral function is a sharp function with most of the spectral weight concentrated around the pole, giving a well-defined dressed dimer.  As the polaron effects become more pronounced, we observe that the pole broadens significantly, hindering any signature of a dressed dimer state. This indicates that two-body bound states are well-defined when polaronic effects are weaker, whereas as polaron dressing increases, it prevents the emergence of well-defined dimers. In our numerical calculations, we add an imaginary term $i\gamma_X/E_n=0.1$ to the impurity Green's function and use an energy cut-off of $\Lambda/E_n=144$ in the BSE.

\begin{figure}[h]
\centering
\includegraphics[width=.9\columnwidth]{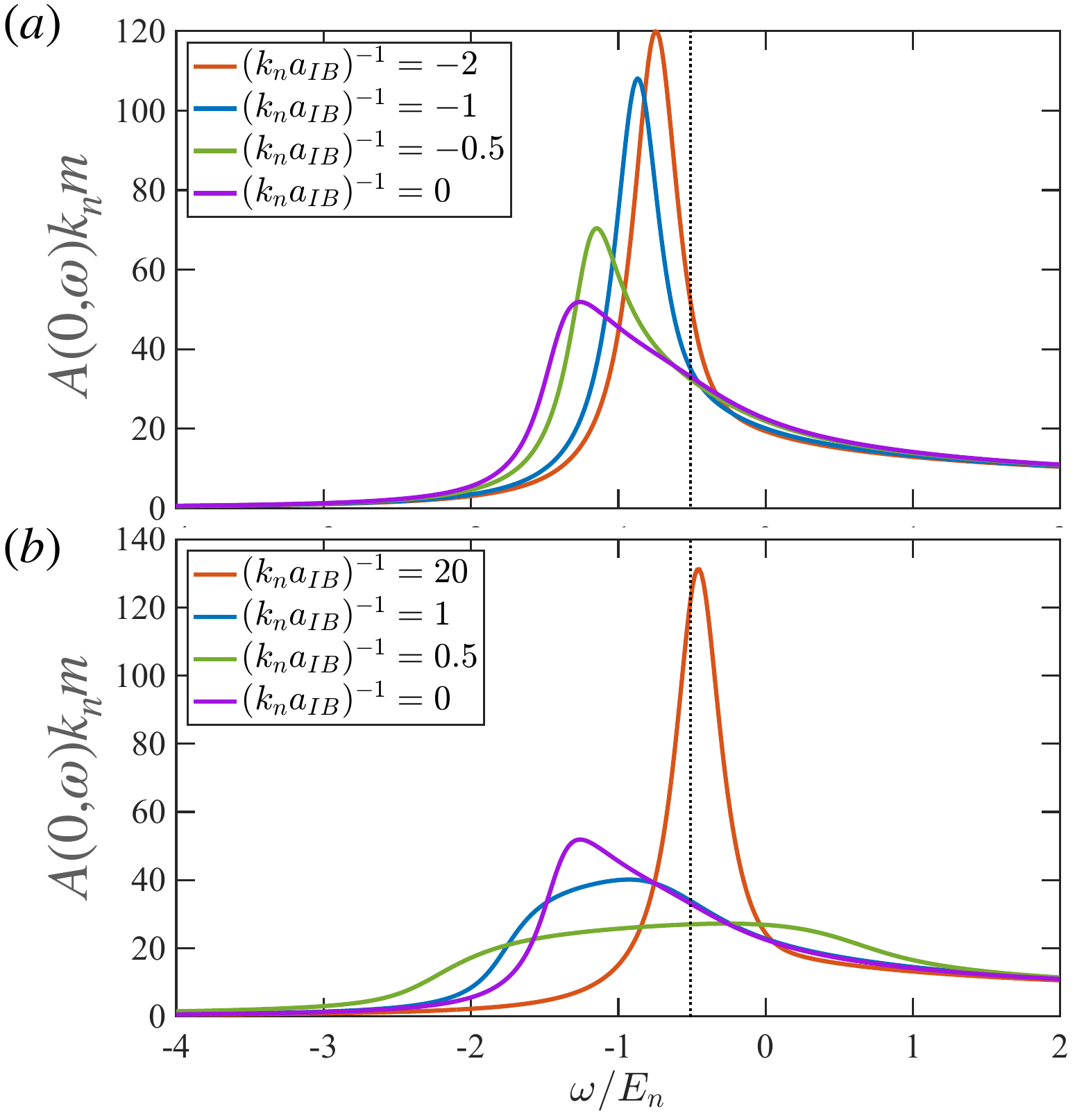}
 \caption{ Spectral function of the impurity-impurity scattering matrix for fixed $1/k_na_{II}=0.5$. (a) Attractive impurity-boson interactions and (b) repulsive impurity-boson interactions. The vertical dashed line indicates the binding energy in the absence of impurity-boson interactions. }
\label{Fig2}
\end{figure}

To further understand the different regimes of dressed dimer formation, we show in Fig.~\ref{Fig2}(a) cross sectioned plots of the spectral function of the scattering matrix for different values of the coupling strength, ranging from weak coupling to the unitary. For weak and intermediate interactions, $\left(k_{n}a_{IB}\right)^{-1}\gg-1$, we observe a well-defined pole in the scattering matrix, appearing as a pronounced peak in the spectral function. This pole is shifted to lower energies due to the attractive impurity interactions, forming the attractive polaron. In the case of strong interactions, $-1<\left(k_{n}a_{IB}\right)^{-1}\le0$, we still observe a pole-like structure in the scattering matrix, but its amplitude decreases, and it becomes asymmetric.

Repulsive impurity-boson interactions lead to contrasting results, as shown in Fig.~\ref{Fig2}(b). For weak repulsive impurity-boson interactions, we observe a pole corresponding to a well-defined bound state, slightly shifted to higher energies, as indicated by the red line for $1/k_n a_{IB} = 20$. Note that for repulsive impurity-boson interactions, a sharp peak is obtained at weaker interactions, compared to attractive impurity-boson interaction as a consequence of the underlying few-body physics for $a_{IB}>0$. For intermediate and strong interactions, the dressed dimer state becomes ill-defined, and the pole structure of the scattering matrix is no longer visible, as shown for $1/k_n a_{IB} = 1$ (blue), $1/k_n a_{IB} = 0.5$ (green), and $1/k_n a_{IB} = 0.0$ (purple). In this regime, the scattering matrix shows a broad feature spanning a wide range of energies. The breakdown of the bound state occurs in a regime where the impurity is formed by two quasiparticle branches, with significant contributions from incoherent excitations beyond the polaron quasiparticle, leading to the broadening of the scattering matrix around the pole. Note that,  for repulsive impurity-boson interactions few-body sates such as trimers, tetramers can be formed~\cite{Astrakharchik2021}.

\begin{figure}[H]
\centering
\includegraphics[width=.9\columnwidth]{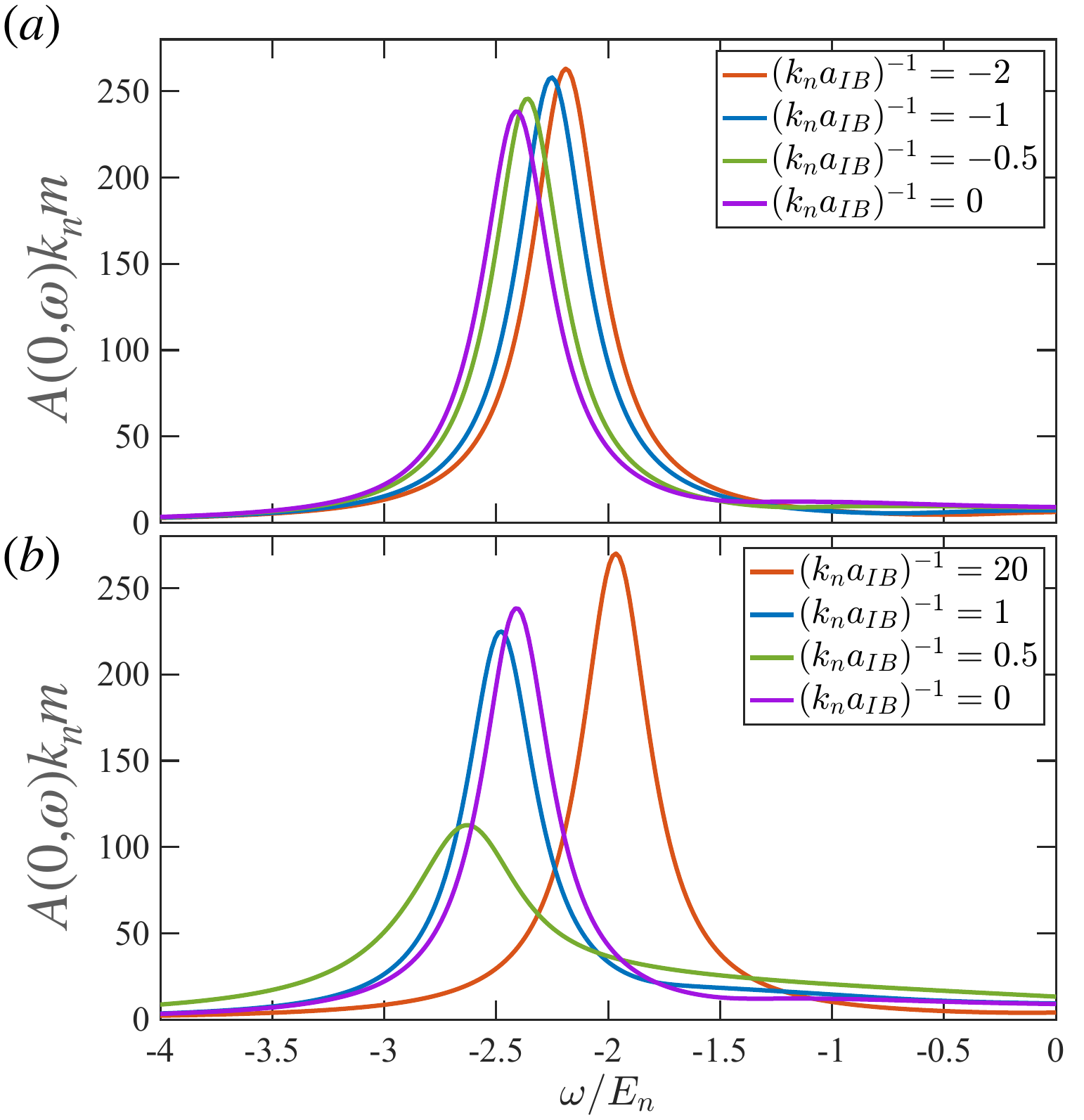}
\caption{ Spectral function of the impurity-impurity scattering matrix for $1/k_na_{II}=1$ (a) Attractive impurity-boson interactions and (b) Repulsive impurity-boson interactions. }
\label{Fig3}
\end{figure}

To further understand how the polaronic dressing influences the formation of the dressed dimer, let us now vary the impurity-impurity scattering length to $1/k_na_{II}=1,$ this case correspond to a vacuum binding energy of   $\varepsilon_I^{(0)}/E_n=-2,$ that is, a more tightly bound state than the discussed in Fig.~\ref{Fig2}. Our results are shown in Fig.~\ref{Fig3} for the same parameters of dimer-boson coupling strength than Fig.~\ref{Fig2} 
 -- following the same color coding. For attractive interactions   Fig.~\ref{Fig3} (a) we observe a clear pole of the scattering matrix which remains well-defined even when the polaron is driven to the unitary regime $1/k_na_{IB}=0.0$. Again, as a consequence of the attractive impurity-boson interaction the pole of the scattering matrix is displaced to negative energies compared to the energy of the bound state in the absence of any impurity-boson interactions. Furthermore, we obtain that the scattering matrix remains much more symmetric for strong interactions than in Fig.~\ref{Fig2}(top).

The persistence of the impurity-impurity bound state is more clearly exhibited for repulsive interactions, in this case, contrary to the results in Fig.~\ref{Fig2} (b) at strong interactions, at $1/k_na_{II}=1,$ we do not find a breakdown of the dimer state. Although the amplitude of the scattering matrix reduces, insights of a bound state can be clearly identified.

\begin{figure}[H]
\centering
\includegraphics[width=.95\columnwidth]{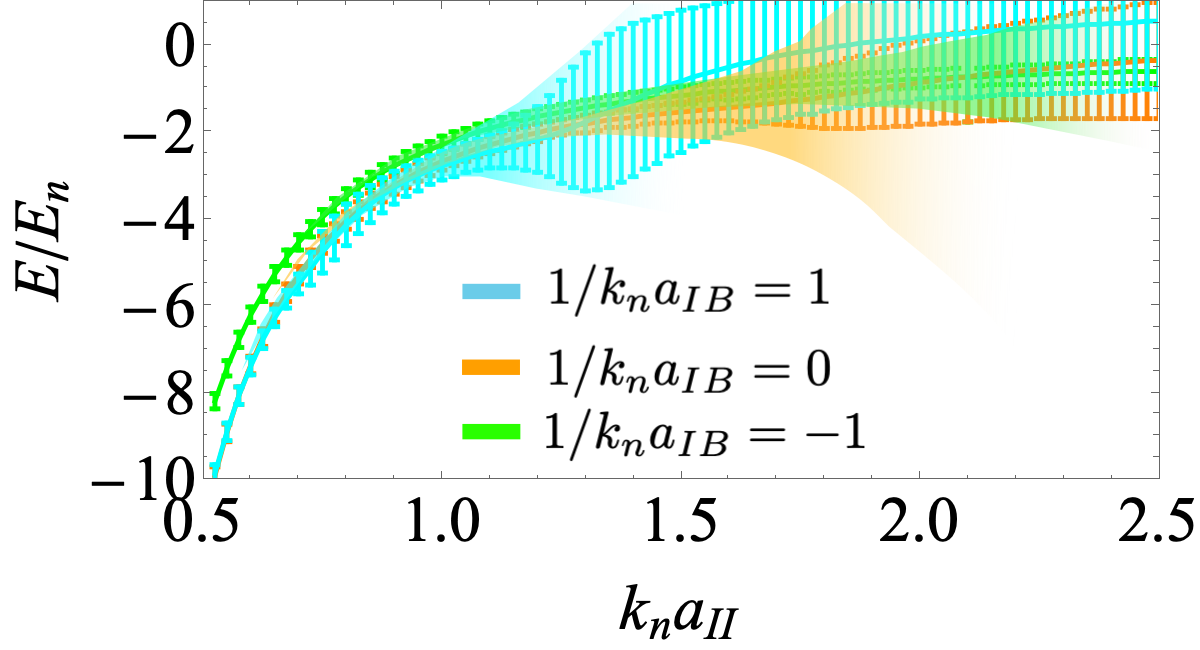}
\caption{  Energy of the bound state as a function of $k_na_{II}$ for several values of the impurity-boson scattering length $k_na_{II},$ the width of the curves is given by the full width at half maximum (FWHM) of the spectral function. The colored shadings are a guide to the eyes to illustrate the regime where the dressed dimer picture breakdown.}
\label{Fig4}
\end{figure}

To understand the breakdown of the dressed dimer we plot in Fig.~\ref{Fig4}  the binding energy of the dressed dimer and its full width at half maximum (FWHM) for intermediate attractive impurity-boson interactions $1/k_na_{IB}=-1.0$  (green curve), strong impurity-boson interaction  $1/k_na_{IB}=0.0$ (orange curve) and for the repulsive branch $1/k_na_{IB}=1.0$ The width of the curves correspond to the FWHM. Figure~\ref{Fig4} allows us to give a very intuitive physical picture to our results: For the polaronic dressing of the dimer, two relevant energies must be compared: the polaron energy $\omega_{\mathbf k}^P$ and the energy of the bare dimer $\varepsilon_I^{(0)}$. First, in the regime where $|\varepsilon_I^{(0)}|\gg \omega_{\mathbf 0}^P$, the polaron effects are weak compared to the binding energy of the molecule. Therefore, the dressed dimer formation remains robust against polaronic dressing. This is evident in all three curves in Fig.~\ref{Fig4}. In these cases, the binding energy of the dressed dimer remains larger than the polaron energy, and the effects of the polaron are barely visible.

On the other hand, in the opposite regime, for $|\varepsilon^{(0)}_I|\ll \omega_{\mathbf 0}^P,$  the dimer is more fragile and sensible to polaron effects, this can be seen in the breakdown of the dimer for the unitary regime where the width of the dimer becomes much larger than the dimer energy, signaling the fading of this bound state. This constructs the phase diagram in Fig.~\ref{Fig0} , where the polaron regime is defined in the limit $|\varepsilon^{(0)}_I|\ll \omega_{\mathbf 0}^P$, whereas the dimer phase is given by $|\varepsilon^{(0)}_I|\gg \omega_{\mathbf 0}^P$. Here, the black dashed line denotes the crossover $|\varepsilon^{(0)}_I|= \omega_{\mathbf 0}^P$ .  For repulsive interactions, we observe a more dramatic interplay between polaron dressing and the dimer as a consequence of the emergence of the repulsive branch, which also leads to the breakdown of the dimer for $|\varepsilon^{(0)}_I|\ll \omega_{\mathbf 0}^P$.  

Our study relies on two approximations: First, the NSCT approximation, which incorporates the impurity-boson Feshbach resonance and is limited to a single Bogoliubov excitation out of the condensate. This formalism~\cite{rath2013field} has proven highly effective in describing Bose polaron experiments~\cite{jorgensen2016observation,hu2016bose,yan2020bose,skou2021non}, found to be in agreement with quantum Monte Carlo calculations~\cite{ardila2019analyzing}, and has been widely used to study 2D Bose and Fermi polarons in semiconductors~\cite{sidler2017fermi,takemura2014polaritonic,Levinsen2019,Bastarrachea2019}. Second, we neglect induced interactions between impurities, which can lead to bipolaron formation~\cite{Naidon2018,Camacho2018a,Camacho2018b}. This approximation holds if the timescale for dimer formation via direct interactions is much shorter than that associated with mediated interactions. Specifically, this remains valid as long as the speed of sound in the BEC,  
$ c_s = \sqrt{4\pi a_{BB} n_0/m_b^2},$  is much smaller than the typical velocity of the dimer, i.e.,  
$1 \gg \frac{c_s}{\sqrt{2|\varepsilon_I^{(0)}|/m_b}} \sim \sqrt{n_0 a_{BB} a_{IB}^2}.$  
Here, \( a_{BB} \) is the boson-boson scattering length. We emphasize that, unlike mediated interactions, which are significantly affected by retardation effects, the {\it direct} two-body interaction is instantaneous.

{\it Conclusions.-} We have studied the polaron effects on the formation of a bound state due to the direct interaction between two impurities. By using a Green's function formalism which allowed us to the treat the single polaron non-perturbatively and by addressing the two-body problem by solving the BSE for the interaction of two impurity atoms we have demonstrated a subtle interplay between polaron physics and an underlying two-body bound state. 
 
 Our results show that for weakly bound dimers, polaron effects can be strong enough to completely break the formation of a dimer whereas tightly bound dimers are more robust towards polaron in the strongly interacting impurity-boson regime.  Our theoretical proposal can be tested with the current state-of-the-art experimental capabilities to probe Feshbach molecules~\cite{Chin2010}. An interesting avenue is to explore beyond the NSCT approximation the competition between polaron-few-body states~\cite{Astrakharchik2021} and the formation of dressed dimers.
 
The study of the impact of beyond Fr\"ohlich polaron dressing on two-body molecular states and bipolaron formation is potentially relevant to the study of Hubbard-Holstein models~\cite{Wellein1996,Trugman2000,bonca2001,Macridin2004}  or organic semiconductors. 

\section{End Matter}

\begin{figure}[H]
\centering
\includegraphics[width=.95\columnwidth]{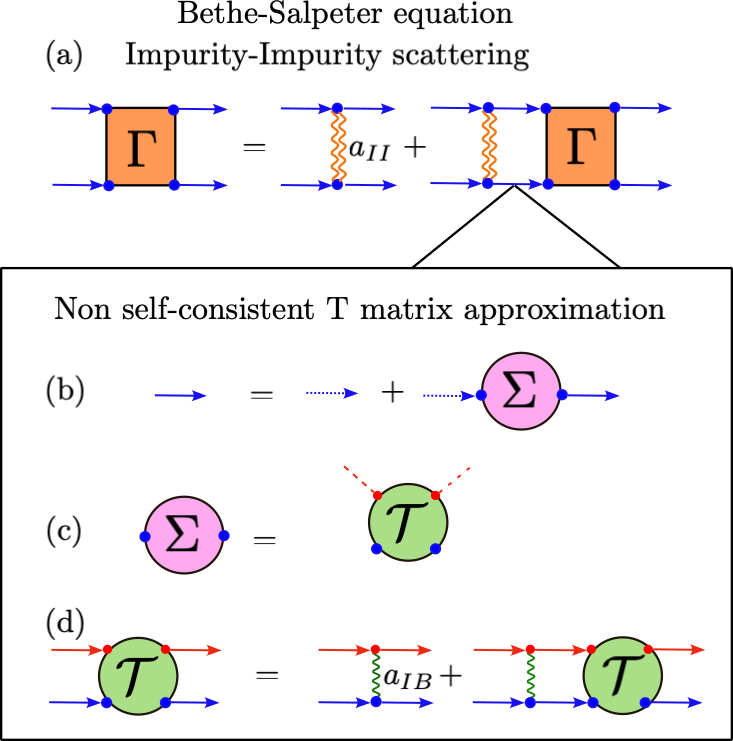}
\caption{(a) In orange, the BSE equation for the impurity-impurity scattering, the bare impurity-impurity interaction is depicted by the double wavy orange line and the impurity Green's function by the solid blue line. (b) The Dyson's equation for impurity Green's function, the dashed line gives the ideal impurity Green's function and the pink circle represents the self-energy, which is calculated following the NSCT approximation as in (b). The red dashed line correspond to condensate bosons whereas the solid red lines depict the propagator of the majority bosons. (d) Ladder approximation for the impurity-boson scattering.  }
\label{FigBSE}
\end{figure}

We can illustrate our formalism in a suitable diagrammatic representation as in Fig.~\ref{FigBSE}. In Fig.~\ref{FigBSE}(a) we show the Bethe-Salpeter equation for the impurity-impurity scattering, which consists of the repeated impurity-impurity scattering with the direct interaction $a_{IB},$ depicted by the double wavy orange line.  To solve the BSE, we obtain the impurity Green's function propagator invoking a non self-consistent T-matrix approximation to describe the dressing of the impurities with the particles of the bath. 

In Fig.~\ref{FigBSE}(b) we represent diagrammatically the Dyson's equation for the impurity propagator (solid blue line), which can be written in terms of the bare impurity propagator (dashed blue line) and the self-energy (pink circle), given by
\begin{gather}
 G_{cc}(\mathbf p,\omega)=G^{(0)}_{cc}(\mathbf p,\omega)+G^{(0)}_{cc}(\mathbf p,\omega)\Sigma_{cc} (\mathbf p,\omega)G_{cc}(\mathbf p,\omega),  
\end{gather}
with 
$$G^{(0)}_{cc}(\mathbf p,\omega)=\frac{1}{\omega-\epsilon_{\mathbf p}^{(c)}},$$
the ideal Green's function for the impurity.

The self-energy is calculated under the NSCT approximation, and is represented in Fig.~\ref{FigBSE}(c), here the red dashed lines correspond to condensate bosons. 
\begin{gather}
 \Sigma_{cc}(\mathbf p,\omega)=n_0\mathcal T(\mathbf p,\omega).   
\end{gather}

Finally, in Fig.~\ref{FigBSE}(d) we show the diagrammatic representation of the T-matrix, giving the ladder approximation for the boson-impurity scattering $a_{IB}$ which we illustrate by the single wavy green line.  For $a_{BB}=0,$ that is an ideal BEC, the impurity-boson T-matrix acquires a simple form:
\begin{gather}
\mathcal T(\mathbf p,\omega)=\frac{2\pi a_{IB}}{m_r}\frac{1}{1+\frac{i2\pi a_{IB}m_r^{3/2}}{m_r\sqrt{2}\pi}\sqrt{\omega-\frac{k^2}{2M}}}.    
\end{gather}

One should note the similarities between the BSE and the T-matrix for impurity-impurity and impurity-boson scattering, respectively. Both account for the two-body physics at different levels: in the former, it provides the exact solution to the two-body problem; in the latter, it captures the Feshbach physics between the impurity and majority bosons. The NSCT provides the impurity Green's function accounting for a non-perturbative dressing of the impurity with the bosons. The NSCT gives the input to solve the BSE for the impurity-impurity interaction retaining the full complexity of the scattering problem.

 {\it Acknowledgments.-} We thank Pietro A. Massignan for the critical reading of our manuscript. A.C.G. acknowledges financial support from UNAM DGAPA PAPIIT Grant No. IA101923, and UNAM DGAPA PAPIME Grants No. PE100924 and No. PIIF23 and Project CONAHCYT No. CBF2023-2024-1765. and L.A.P.A acknowledges financial support from PNRR MUR project PE0000023-NQSTI.

The data supporting the findings of this study are openly available below ~\cite{camacho_guardian_2024_14516566}.

\bibliography{references}

\begin{thebibliography}{92}%
\makeatletter
\providecommand \@ifxundefined [1]{%
 \@ifx{#1\undefined}
}%
\providecommand \@ifnum [1]{%
 \ifnum #1\expandafter \@firstoftwo
 \else \expandafter \@secondoftwo
 \fi
}%
\providecommand \@ifx [1]{%
 \ifx #1\expandafter \@firstoftwo
 \else \expandafter \@secondoftwo
 \fi
}%
\providecommand \natexlab [1]{#1}%
\providecommand \enquote  [1]{``#1''}%
\providecommand \bibnamefont  [1]{#1}%
\providecommand \bibfnamefont [1]{#1}%
\providecommand \citenamefont [1]{#1}%
\providecommand \href@noop [0]{\@secondoftwo}%
\providecommand \href [0]{\begingroup \@sanitize@url \@href}%
\providecommand \@href[1]{\@@startlink{#1}\@@href}%
\providecommand \@@href[1]{\endgroup#1\@@endlink}%
\providecommand \@sanitize@url [0]{\catcode `\\12\catcode `\$12\catcode
  `\&12\catcode `\#12\catcode `\^12\catcode `\_12\catcode `\%12\relax}%
\providecommand \@@startlink[1]{}%
\providecommand \@@endlink[0]{}%
\providecommand \url  [0]{\begingroup\@sanitize@url \@url }%
\providecommand \@url [1]{\endgroup\@href {#1}{\urlprefix }}%
\providecommand \urlprefix  [0]{URL }%
\providecommand \Eprint [0]{\href }%
\providecommand \doibase [0]{http://dx.doi.org/}%
\providecommand \selectlanguage [0]{\@gobble}%
\providecommand \bibinfo  [0]{\@secondoftwo}%
\providecommand \bibfield  [0]{\@secondoftwo}%
\providecommand \translation [1]{[#1]}%
\providecommand \BibitemOpen [0]{}%
\providecommand \bibitemStop [0]{}%
\providecommand \bibitemNoStop [0]{.\EOS\space}%
\providecommand \EOS [0]{\spacefactor3000\relax}%
\providecommand \BibitemShut  [1]{\csname bibitem#1\endcsname}%
\let\auto@bib@innerbib\@empty
\bibitem [{\citenamefont {Landau}\ and\ \citenamefont
  {Pekar}(1948)}]{landau1948effective}%
  \BibitemOpen
  \bibfield  {author} {\bibinfo {author} {\bibfnamefont {LD}~\bibnamefont
  {Landau}}\ and\ \bibinfo {author} {\bibfnamefont {SI}~\bibnamefont {Pekar}},\
  }\bibfield  {title} {\enquote {\bibinfo {title} {Effective mass of a
  polaron},}\ }\href@noop {} {\bibfield  {journal} {\bibinfo  {journal} {Zh.
  Eksp. Teor. Fiz}\ }\textbf {\bibinfo {volume} {18}},\ \bibinfo {pages}
  {419--423} (\bibinfo {year} {1948})}\BibitemShut {NoStop}%
\bibitem [{\citenamefont {J{\o}rgensen}\ \emph {et~al.}(2016)\citenamefont
  {J{\o}rgensen}, \citenamefont {Wacker}, \citenamefont {Skalmstang},
  \citenamefont {Parish}, \citenamefont {Levinsen}, \citenamefont
  {Christensen}, \citenamefont {Bruun},\ and\ \citenamefont
  {Arlt}}]{jorgensen2016observation}%
  \BibitemOpen
  \bibfield  {author} {\bibinfo {author} {\bibfnamefont {Nils~B}\ \bibnamefont
  {J{\o}rgensen}}, \bibinfo {author} {\bibfnamefont {Lars}\ \bibnamefont
  {Wacker}}, \bibinfo {author} {\bibfnamefont {Kristoffer~T}\ \bibnamefont
  {Skalmstang}}, \bibinfo {author} {\bibfnamefont {Meera~M}\ \bibnamefont
  {Parish}}, \bibinfo {author} {\bibfnamefont {Jesper}\ \bibnamefont
  {Levinsen}}, \bibinfo {author} {\bibfnamefont {Rasmus~S}\ \bibnamefont
  {Christensen}}, \bibinfo {author} {\bibfnamefont {Georg~M}\ \bibnamefont
  {Bruun}}, \ and\ \bibinfo {author} {\bibfnamefont {Jan~J}\ \bibnamefont
  {Arlt}},\ }\bibfield  {title} {\enquote {\bibinfo {title} {Observation of
  attractive and repulsive polarons in a bose-einstein condensate},}\
  }\href@noop {} {\bibfield  {journal} {\bibinfo  {journal} {Physical review
  letters}\ }\textbf {\bibinfo {volume} {117}},\ \bibinfo {pages} {055302}
  (\bibinfo {year} {2016})}\BibitemShut {NoStop}%
\bibitem [{\citenamefont {Hu}\ \emph {et~al.}(2016)\citenamefont {Hu},
  \citenamefont {Van~de Graaff}, \citenamefont {Kedar}, \citenamefont {Corson},
  \citenamefont {Cornell},\ and\ \citenamefont {Jin}}]{hu2016bose}%
  \BibitemOpen
  \bibfield  {author} {\bibinfo {author} {\bibfnamefont {Ming-Guang}\
  \bibnamefont {Hu}}, \bibinfo {author} {\bibfnamefont {Michael~J}\
  \bibnamefont {Van~de Graaff}}, \bibinfo {author} {\bibfnamefont {Dhruv}\
  \bibnamefont {Kedar}}, \bibinfo {author} {\bibfnamefont {John~P}\
  \bibnamefont {Corson}}, \bibinfo {author} {\bibfnamefont {Eric~A}\
  \bibnamefont {Cornell}}, \ and\ \bibinfo {author} {\bibfnamefont {Deborah~S}\
  \bibnamefont {Jin}},\ }\bibfield  {title} {\enquote {\bibinfo {title} {Bose
  polarons in the strongly interacting regime},}\ }\href@noop {} {\bibfield
  {journal} {\bibinfo  {journal} {Physical review letters}\ }\textbf {\bibinfo
  {volume} {117}},\ \bibinfo {pages} {055301} (\bibinfo {year}
  {2016})}\BibitemShut {NoStop}%
\bibitem [{\citenamefont {Pekar}(1969)}]{pekar1969theory}%
  \BibitemOpen
  \bibfield  {author} {\bibinfo {author} {\bibfnamefont {SI}~\bibnamefont
  {Pekar}},\ }\bibfield  {title} {\enquote {\bibinfo {title} {Theory of
  polarons in many-valley crystals i. weak interaction between electron and
  lattice polarization field},}\ }\href@noop {} {\bibfield  {journal} {\bibinfo
   {journal} {Soviet Physics Jetp}\ }\textbf {\bibinfo {volume} {28}} (\bibinfo
  {year} {1969})}\BibitemShut {NoStop}%
\bibitem [{\citenamefont {Eagles}(1963)}]{Eagles1963}%
  \BibitemOpen
  \bibfield  {author} {\bibinfo {author} {\bibfnamefont {D.~M.}\ \bibnamefont
  {Eagles}},\ }\bibfield  {title} {\enquote {\bibinfo {title} {Optical
  absorption in ionic crystals involving small polarons},}\ }\href {\doibase
  10.1103/PhysRev.130.1381} {\bibfield  {journal} {\bibinfo  {journal} {Phys.
  Rev.}\ }\textbf {\bibinfo {volume} {130}},\ \bibinfo {pages} {1381--1400}
  (\bibinfo {year} {1963})}\BibitemShut {NoStop}%
\bibitem [{\citenamefont {Devreese}\ and\ \citenamefont
  {Alexandrov}(2009)}]{devreese2009frohlich}%
  \BibitemOpen
  \bibfield  {author} {\bibinfo {author} {\bibfnamefont {Jozef~T}\ \bibnamefont
  {Devreese}}\ and\ \bibinfo {author} {\bibfnamefont {Alexandre~S}\
  \bibnamefont {Alexandrov}},\ }\bibfield  {title} {\enquote {\bibinfo {title}
  {Fr{\"o}hlich polaron and bipolaron: recent developments},}\ }\href@noop {}
  {\bibfield  {journal} {\bibinfo  {journal} {Reports on Progress in Physics}\
  }\textbf {\bibinfo {volume} {72}},\ \bibinfo {pages} {066501} (\bibinfo
  {year} {2009})}\BibitemShut {NoStop}%
\bibitem [{\citenamefont {Grusdt}\ and\ \citenamefont
  {Fleischhauer}(2016)}]{Grusdt2014}%
  \BibitemOpen
  \bibfield  {author} {\bibinfo {author} {\bibfnamefont {Fabian}\ \bibnamefont
  {Grusdt}}\ and\ \bibinfo {author} {\bibfnamefont {Michael}\ \bibnamefont
  {Fleischhauer}},\ }\bibfield  {title} {\enquote {\bibinfo {title} {Tunable
  polarons of slow-light polaritons in a two-dimensional bose-einstein
  condensate},}\ }\href {\doibase 10.1103/PhysRevLett.116.053602} {\bibfield
  {journal} {\bibinfo  {journal} {Phys. Rev. Lett.}\ }\textbf {\bibinfo
  {volume} {116}},\ \bibinfo {pages} {053602} (\bibinfo {year}
  {2016})}\BibitemShut {NoStop}%
\bibitem [{\citenamefont {Nielsen}\ \emph {et~al.}(2020)\citenamefont
  {Nielsen}, \citenamefont {Camacho-Guardian}, \citenamefont {Bruun},\ and\
  \citenamefont {Pohl}}]{Nielsen2020}%
  \BibitemOpen
  \bibfield  {author} {\bibinfo {author} {\bibfnamefont {K.~Knakkergaard}\
  \bibnamefont {Nielsen}}, \bibinfo {author} {\bibfnamefont {A.}~\bibnamefont
  {Camacho-Guardian}}, \bibinfo {author} {\bibfnamefont {G.~M.}\ \bibnamefont
  {Bruun}}, \ and\ \bibinfo {author} {\bibfnamefont {T.}~\bibnamefont {Pohl}},\
  }\bibfield  {title} {\enquote {\bibinfo {title} {Superfluid flow of polaron
  polaritons above landau's critical velocity},}\ }\href {\doibase
  10.1103/PhysRevLett.125.035301} {\bibfield  {journal} {\bibinfo  {journal}
  {Phys. Rev. Lett.}\ }\textbf {\bibinfo {volume} {125}},\ \bibinfo {pages}
  {035301} (\bibinfo {year} {2020})}\BibitemShut {NoStop}%
\bibitem [{\citenamefont {Camacho-Guardian}\ \emph {et~al.}(2020)\citenamefont
  {Camacho-Guardian}, \citenamefont {Nielsen}, \citenamefont {Pohl},\ and\
  \citenamefont {Bruun}}]{Camacho2020}%
  \BibitemOpen
  \bibfield  {author} {\bibinfo {author} {\bibfnamefont {A.}~\bibnamefont
  {Camacho-Guardian}}, \bibinfo {author} {\bibfnamefont {K.~Knakkergaard}\
  \bibnamefont {Nielsen}}, \bibinfo {author} {\bibfnamefont {T.}~\bibnamefont
  {Pohl}}, \ and\ \bibinfo {author} {\bibfnamefont {G.~M.}\ \bibnamefont
  {Bruun}},\ }\bibfield  {title} {\enquote {\bibinfo {title} {Polariton
  dynamics in strongly interacting quantum many-body systems},}\ }\href
  {\doibase 10.1103/PhysRevResearch.2.023102} {\bibfield  {journal} {\bibinfo
  {journal} {Phys. Rev. Research}\ }\textbf {\bibinfo {volume} {2}},\ \bibinfo
  {pages} {023102} (\bibinfo {year} {2020})}\BibitemShut {NoStop}%
\bibitem [{\citenamefont {Srimath~Kandada}\ and\ \citenamefont
  {Silva}(2020)}]{srimath2020exciton}%
  \BibitemOpen
  \bibfield  {author} {\bibinfo {author} {\bibfnamefont {Ajay~Ram}\
  \bibnamefont {Srimath~Kandada}}\ and\ \bibinfo {author} {\bibfnamefont
  {Carlos}\ \bibnamefont {Silva}},\ }\bibfield  {title} {\enquote {\bibinfo
  {title} {Exciton polarons in two-dimensional hybrid metal-halide
  perovskites},}\ }\href@noop {} {\bibfield  {journal} {\bibinfo  {journal}
  {The Journal of Physical Chemistry Letters}\ }\textbf {\bibinfo {volume}
  {11}},\ \bibinfo {pages} {3173--3184} (\bibinfo {year} {2020})}\BibitemShut
  {NoStop}%
\bibitem [{\citenamefont {Franchini}\ \emph {et~al.}(2021)\citenamefont
  {Franchini}, \citenamefont {Reticcioli}, \citenamefont {Setvin},\ and\
  \citenamefont {Diebold}}]{franchini2021polarons}%
  \BibitemOpen
  \bibfield  {author} {\bibinfo {author} {\bibfnamefont {Cesare}\ \bibnamefont
  {Franchini}}, \bibinfo {author} {\bibfnamefont {Michele}\ \bibnamefont
  {Reticcioli}}, \bibinfo {author} {\bibfnamefont {Martin}\ \bibnamefont
  {Setvin}}, \ and\ \bibinfo {author} {\bibfnamefont {Ulrike}\ \bibnamefont
  {Diebold}},\ }\bibfield  {title} {\enquote {\bibinfo {title} {Polarons in
  materials},}\ }\href@noop {} {\bibfield  {journal} {\bibinfo  {journal}
  {Nature Reviews Materials}\ }\textbf {\bibinfo {volume} {6}},\ \bibinfo
  {pages} {560--586} (\bibinfo {year} {2021})}\BibitemShut {NoStop}%
\bibitem [{\citenamefont {Tao}\ \emph {et~al.}(2021)\citenamefont {Tao},
  \citenamefont {Zhang}, \citenamefont {Zhou}, \citenamefont {Zhao},\ and\
  \citenamefont {Zhu}}]{tao2021momentarily}%
  \BibitemOpen
  \bibfield  {author} {\bibinfo {author} {\bibfnamefont {Weijian}\ \bibnamefont
  {Tao}}, \bibinfo {author} {\bibfnamefont {Chi}\ \bibnamefont {Zhang}},
  \bibinfo {author} {\bibfnamefont {Qiaohui}\ \bibnamefont {Zhou}}, \bibinfo
  {author} {\bibfnamefont {Yida}\ \bibnamefont {Zhao}}, \ and\ \bibinfo
  {author} {\bibfnamefont {Haiming}\ \bibnamefont {Zhu}},\ }\bibfield  {title}
  {\enquote {\bibinfo {title} {Momentarily trapped exciton polaron in
  two-dimensional lead halide perovskites},}\ }\href@noop {} {\bibfield
  {journal} {\bibinfo  {journal} {Nature Communications}\ }\textbf {\bibinfo
  {volume} {12}},\ \bibinfo {pages} {1400} (\bibinfo {year}
  {2021})}\BibitemShut {NoStop}%
\bibitem [{\citenamefont {Bourelle}\ \emph {et~al.}(2022)\citenamefont
  {Bourelle}, \citenamefont {Camargo}, \citenamefont {Ghosh}, \citenamefont
  {Neumann}, \citenamefont {van~de Goor}, \citenamefont {Shivanna},
  \citenamefont {Winkler}, \citenamefont {Cerullo},\ and\ \citenamefont
  {Deschler}}]{bourelle2022optical}%
  \BibitemOpen
  \bibfield  {author} {\bibinfo {author} {\bibfnamefont {Sean~A}\ \bibnamefont
  {Bourelle}}, \bibinfo {author} {\bibfnamefont {Franco~VA}\ \bibnamefont
  {Camargo}}, \bibinfo {author} {\bibfnamefont {Soumen}\ \bibnamefont {Ghosh}},
  \bibinfo {author} {\bibfnamefont {Timo}\ \bibnamefont {Neumann}}, \bibinfo
  {author} {\bibfnamefont {Tim~WJ}\ \bibnamefont {van~de Goor}}, \bibinfo
  {author} {\bibfnamefont {Ravichandran}\ \bibnamefont {Shivanna}}, \bibinfo
  {author} {\bibfnamefont {Thomas}\ \bibnamefont {Winkler}}, \bibinfo {author}
  {\bibfnamefont {Giulio}\ \bibnamefont {Cerullo}}, \ and\ \bibinfo {author}
  {\bibfnamefont {Felix}\ \bibnamefont {Deschler}},\ }\bibfield  {title}
  {\enquote {\bibinfo {title} {Optical control of exciton spin dynamics in
  layered metal halide perovskites via polaronic state formation},}\
  }\href@noop {} {\bibfield  {journal} {\bibinfo  {journal} {Nature
  Communications}\ }\textbf {\bibinfo {volume} {13}},\ \bibinfo {pages} {3320}
  (\bibinfo {year} {2022})}\BibitemShut {NoStop}%
\bibitem [{\citenamefont {Kobyakov}\ and\ \citenamefont
  {Pethick}(2016)}]{Kobyakov2016}%
  \BibitemOpen
  \bibfield  {author} {\bibinfo {author} {\bibfnamefont {D.}~\bibnamefont
  {Kobyakov}}\ and\ \bibinfo {author} {\bibfnamefont {C.~J.}\ \bibnamefont
  {Pethick}},\ }\bibfield  {title} {\enquote {\bibinfo {title} {Nucleus-nucleus
  interactions in the inner crust of neutron stars},}\ }\href {\doibase
  10.1103/PhysRevC.94.055806} {\bibfield  {journal} {\bibinfo  {journal} {Phys.
  Rev. C}\ }\textbf {\bibinfo {volume} {94}},\ \bibinfo {pages} {055806}
  (\bibinfo {year} {2016})}\BibitemShut {NoStop}%
\bibitem [{\citenamefont {Tajima}\ \emph {et~al.}(2024)\citenamefont {Tajima},
  \citenamefont {Moriya}, \citenamefont {Horiuchi}, \citenamefont {Nakano},\
  and\ \citenamefont {Iida}}]{tajima2024polaronic}%
  \BibitemOpen
  \bibfield  {author} {\bibinfo {author} {\bibfnamefont {Hiroyuki}\
  \bibnamefont {Tajima}}, \bibinfo {author} {\bibfnamefont {Hajime}\
  \bibnamefont {Moriya}}, \bibinfo {author} {\bibfnamefont {Wataru}\
  \bibnamefont {Horiuchi}}, \bibinfo {author} {\bibfnamefont {Eiji}\
  \bibnamefont {Nakano}}, \ and\ \bibinfo {author} {\bibfnamefont {Kei}\
  \bibnamefont {Iida}},\ }\bibfield  {title} {\enquote {\bibinfo {title}
  {Polaronic proton and diproton clustering in neutron-rich matter},}\
  }\href@noop {} {\bibfield  {journal} {\bibinfo  {journal} {Physics Letters
  B}\ }\textbf {\bibinfo {volume} {851}},\ \bibinfo {pages} {138567} (\bibinfo
  {year} {2024})}\BibitemShut {NoStop}%
\bibitem [{\citenamefont {Ardila}\ \emph {et~al.}(2019)\citenamefont {Ardila},
  \citenamefont {J{\o}rgensen}, \citenamefont {Pohl}, \citenamefont {Giorgini},
  \citenamefont {Bruun},\ and\ \citenamefont {Arlt}}]{ardila2019analyzing}%
  \BibitemOpen
  \bibfield  {author} {\bibinfo {author} {\bibfnamefont {LA~Pe{\~n}a}\
  \bibnamefont {Ardila}}, \bibinfo {author} {\bibfnamefont {NB}~\bibnamefont
  {J{\o}rgensen}}, \bibinfo {author} {\bibfnamefont {T}~\bibnamefont {Pohl}},
  \bibinfo {author} {\bibfnamefont {S}~\bibnamefont {Giorgini}}, \bibinfo
  {author} {\bibfnamefont {GM}~\bibnamefont {Bruun}}, \ and\ \bibinfo {author}
  {\bibfnamefont {JJ}~\bibnamefont {Arlt}},\ }\bibfield  {title} {\enquote
  {\bibinfo {title} {Analyzing a bose polaron across resonant interactions},}\
  }\href@noop {} {\bibfield  {journal} {\bibinfo  {journal} {Physical Review
  A}\ }\textbf {\bibinfo {volume} {99}},\ \bibinfo {pages} {063607} (\bibinfo
  {year} {2019})}\BibitemShut {NoStop}%
\bibitem [{\citenamefont {Yan}\ \emph {et~al.}(2020)\citenamefont {Yan},
  \citenamefont {Ni}, \citenamefont {Robens},\ and\ \citenamefont
  {Zwierlein}}]{yan2020bose}%
  \BibitemOpen
  \bibfield  {author} {\bibinfo {author} {\bibfnamefont {Zoe~Z}\ \bibnamefont
  {Yan}}, \bibinfo {author} {\bibfnamefont {Yiqi}\ \bibnamefont {Ni}}, \bibinfo
  {author} {\bibfnamefont {Carsten}\ \bibnamefont {Robens}}, \ and\ \bibinfo
  {author} {\bibfnamefont {Martin~W}\ \bibnamefont {Zwierlein}},\ }\bibfield
  {title} {\enquote {\bibinfo {title} {Bose polarons near quantum
  criticality},}\ }\href@noop {} {\bibfield  {journal} {\bibinfo  {journal}
  {Science}\ }\textbf {\bibinfo {volume} {368}},\ \bibinfo {pages} {190--194}
  (\bibinfo {year} {2020})}\BibitemShut {NoStop}%
\bibitem [{\citenamefont {Skou}\ \emph
  {et~al.}(2021{\natexlab{a}})\citenamefont {Skou}, \citenamefont {Skov},
  \citenamefont {J{\o}rgensen}, \citenamefont {Nielsen}, \citenamefont
  {Camacho-Guardian}, \citenamefont {Pohl}, \citenamefont {Bruun},\ and\
  \citenamefont {Arlt}}]{skou2021non}%
  \BibitemOpen
  \bibfield  {author} {\bibinfo {author} {\bibfnamefont {Magnus~G}\
  \bibnamefont {Skou}}, \bibinfo {author} {\bibfnamefont {Thomas~G}\
  \bibnamefont {Skov}}, \bibinfo {author} {\bibfnamefont {Nils~B}\ \bibnamefont
  {J{\o}rgensen}}, \bibinfo {author} {\bibfnamefont {Kristian~K}\ \bibnamefont
  {Nielsen}}, \bibinfo {author} {\bibfnamefont {Arturo}\ \bibnamefont
  {Camacho-Guardian}}, \bibinfo {author} {\bibfnamefont {Thomas}\ \bibnamefont
  {Pohl}}, \bibinfo {author} {\bibfnamefont {Georg~M}\ \bibnamefont {Bruun}}, \
  and\ \bibinfo {author} {\bibfnamefont {Jan~J}\ \bibnamefont {Arlt}},\
  }\bibfield  {title} {\enquote {\bibinfo {title} {Non-equilibrium quantum
  dynamics and formation of the bose polaron},}\ }\href@noop {} {\bibfield
  {journal} {\bibinfo  {journal} {Nature Physics}\ }\textbf {\bibinfo {volume}
  {17}},\ \bibinfo {pages} {731--735} (\bibinfo {year}
  {2021}{\natexlab{a}})}\BibitemShut {NoStop}%
\bibitem [{\citenamefont {Skou}\ \emph {et~al.}(2022)\citenamefont {Skou},
  \citenamefont {Nielsen}, \citenamefont {Skov}, \citenamefont {Morgen},
  \citenamefont {J\o{}rgensen}, \citenamefont {Camacho-Guardian}, \citenamefont
  {Pohl}, \citenamefont {Bruun},\ and\ \citenamefont {Arlt}}]{Skou2022}%
  \BibitemOpen
  \bibfield  {author} {\bibinfo {author} {\bibfnamefont {Magnus~G.}\
  \bibnamefont {Skou}}, \bibinfo {author} {\bibfnamefont {Kristian~K.}\
  \bibnamefont {Nielsen}}, \bibinfo {author} {\bibfnamefont {Thomas~G.}\
  \bibnamefont {Skov}}, \bibinfo {author} {\bibfnamefont {Andreas~M.}\
  \bibnamefont {Morgen}}, \bibinfo {author} {\bibfnamefont {Nils~B.}\
  \bibnamefont {J\o{}rgensen}}, \bibinfo {author} {\bibfnamefont {Arturo}\
  \bibnamefont {Camacho-Guardian}}, \bibinfo {author} {\bibfnamefont {Thomas}\
  \bibnamefont {Pohl}}, \bibinfo {author} {\bibfnamefont {Georg~M.}\
  \bibnamefont {Bruun}}, \ and\ \bibinfo {author} {\bibfnamefont {Jan~J.}\
  \bibnamefont {Arlt}},\ }\bibfield  {title} {\enquote {\bibinfo {title} {Life
  and death of the bose polaron},}\ }\href {\doibase
  10.1103/PhysRevResearch.4.043093} {\bibfield  {journal} {\bibinfo  {journal}
  {Phys. Rev. Res.}\ }\textbf {\bibinfo {volume} {4}},\ \bibinfo {pages}
  {043093} (\bibinfo {year} {2022})}\BibitemShut {NoStop}%
\bibitem [{\citenamefont {Etrych}\ \emph {et~al.}(2024)\citenamefont {Etrych},
  \citenamefont {Martirosyan}, \citenamefont {Cao}, \citenamefont {Ho},
  \citenamefont {Hadzibabic},\ and\ \citenamefont
  {Eigen}}]{etrych2024universal}%
  \BibitemOpen
  \bibfield  {author} {\bibinfo {author} {\bibfnamefont {Ji{\v{r}}{\'\i}}\
  \bibnamefont {Etrych}}, \bibinfo {author} {\bibfnamefont {Gevorg}\
  \bibnamefont {Martirosyan}}, \bibinfo {author} {\bibfnamefont {Alec}\
  \bibnamefont {Cao}}, \bibinfo {author} {\bibfnamefont {Christopher~J}\
  \bibnamefont {Ho}}, \bibinfo {author} {\bibfnamefont {Zoran}\ \bibnamefont
  {Hadzibabic}}, \ and\ \bibinfo {author} {\bibfnamefont {Christoph}\
  \bibnamefont {Eigen}},\ }\bibfield  {title} {\enquote {\bibinfo {title}
  {Universal quantum dynamics of bose polarons},}\ }\href@noop {} {\bibfield
  {journal} {\bibinfo  {journal} {arXiv preprint arXiv:2402.14816}\ } (\bibinfo
  {year} {2024})}\BibitemShut {NoStop}%
\bibitem [{\citenamefont {Rath}\ and\ \citenamefont
  {Schmidt}(2013)}]{rath2013field}%
  \BibitemOpen
  \bibfield  {author} {\bibinfo {author} {\bibfnamefont {Steffen~Patrick}\
  \bibnamefont {Rath}}\ and\ \bibinfo {author} {\bibfnamefont {Richard}\
  \bibnamefont {Schmidt}},\ }\bibfield  {title} {\enquote {\bibinfo {title}
  {Field-theoretical study of the bose polaron},}\ }\href@noop {} {\bibfield
  {journal} {\bibinfo  {journal} {Physical Review A}\ }\textbf {\bibinfo
  {volume} {88}},\ \bibinfo {pages} {053632} (\bibinfo {year}
  {2013})}\BibitemShut {NoStop}%
\bibitem [{\citenamefont {Ardila}\ and\ \citenamefont
  {Giorgini}(2015)}]{Ardila2015impurity}%
  \BibitemOpen
  \bibfield  {author} {\bibinfo {author} {\bibfnamefont {LA~Pe{\~n}a}\
  \bibnamefont {Ardila}}\ and\ \bibinfo {author} {\bibfnamefont
  {S}~\bibnamefont {Giorgini}},\ }\bibfield  {title} {\enquote {\bibinfo
  {title} {Impurity in a bose-einstein condensate: Study of the attractive and
  repulsive branch using quantum monte carlo methods},}\ }\href@noop {}
  {\bibfield  {journal} {\bibinfo  {journal} {Physical Review A}\ }\textbf
  {\bibinfo {volume} {92}},\ \bibinfo {pages} {033612} (\bibinfo {year}
  {2015})}\BibitemShut {NoStop}%
\bibitem [{\citenamefont {Shchadilova}\ \emph {et~al.}(2016)\citenamefont
  {Shchadilova}, \citenamefont {Schmidt}, \citenamefont {Grusdt},\ and\
  \citenamefont {Demler}}]{Shchadilova2016}%
  \BibitemOpen
  \bibfield  {author} {\bibinfo {author} {\bibfnamefont {Yulia~E.}\
  \bibnamefont {Shchadilova}}, \bibinfo {author} {\bibfnamefont {Richard}\
  \bibnamefont {Schmidt}}, \bibinfo {author} {\bibfnamefont {Fabian}\
  \bibnamefont {Grusdt}}, \ and\ \bibinfo {author} {\bibfnamefont {Eugene}\
  \bibnamefont {Demler}},\ }\bibfield  {title} {\enquote {\bibinfo {title}
  {Quantum dynamics of ultracold bose polarons},}\ }\href {\doibase
  10.1103/PhysRevLett.117.113002} {\bibfield  {journal} {\bibinfo  {journal}
  {Phys. Rev. Lett.}\ }\textbf {\bibinfo {volume} {117}},\ \bibinfo {pages}
  {113002} (\bibinfo {year} {2016})}\BibitemShut {NoStop}%
\bibitem [{\citenamefont {Grusdt}\ \emph {et~al.}(2018)\citenamefont {Grusdt},
  \citenamefont {Seetharam}, \citenamefont {Shchadilova},\ and\ \citenamefont
  {Demler}}]{grusdt2018strong}%
  \BibitemOpen
  \bibfield  {author} {\bibinfo {author} {\bibfnamefont {Fabian}\ \bibnamefont
  {Grusdt}}, \bibinfo {author} {\bibfnamefont {Kushal}\ \bibnamefont
  {Seetharam}}, \bibinfo {author} {\bibfnamefont {Yulia}\ \bibnamefont
  {Shchadilova}}, \ and\ \bibinfo {author} {\bibfnamefont {Eugene}\
  \bibnamefont {Demler}},\ }\bibfield  {title} {\enquote {\bibinfo {title}
  {Strong-coupling bose polarons out of equilibrium: Dynamical
  renormalization-group approach},}\ }\href@noop {} {\bibfield  {journal}
  {\bibinfo  {journal} {Physical Review A}\ }\textbf {\bibinfo {volume} {97}},\
  \bibinfo {pages} {033612} (\bibinfo {year} {2018})}\BibitemShut {NoStop}%
\bibitem [{\citenamefont {Christensen}\ \emph {et~al.}(2015)\citenamefont
  {Christensen}, \citenamefont {Levinsen},\ and\ \citenamefont
  {Bruun}}]{Christensen2015}%
  \BibitemOpen
  \bibfield  {author} {\bibinfo {author} {\bibfnamefont {Rasmus~S\o{}gaard}\
  \bibnamefont {Christensen}}, \bibinfo {author} {\bibfnamefont {Jesper}\
  \bibnamefont {Levinsen}}, \ and\ \bibinfo {author} {\bibfnamefont {Georg~M.}\
  \bibnamefont {Bruun}},\ }\bibfield  {title} {\enquote {\bibinfo {title}
  {Quasiparticle properties of a mobile impurity in a bose-einstein
  condensate},}\ }\href {\doibase 10.1103/PhysRevLett.115.160401} {\bibfield
  {journal} {\bibinfo  {journal} {Phys. Rev. Lett.}\ }\textbf {\bibinfo
  {volume} {115}},\ \bibinfo {pages} {160401} (\bibinfo {year}
  {2015})}\BibitemShut {NoStop}%
\bibitem [{\citenamefont {Christianen}\ \emph {et~al.}(2023)\citenamefont
  {Christianen}, \citenamefont {Cirac},\ and\ \citenamefont
  {Schmidt}}]{christianen2023phase}%
  \BibitemOpen
  \bibfield  {author} {\bibinfo {author} {\bibfnamefont {Arthur}\ \bibnamefont
  {Christianen}}, \bibinfo {author} {\bibfnamefont {J~Ignacio}\ \bibnamefont
  {Cirac}}, \ and\ \bibinfo {author} {\bibfnamefont {Richard}\ \bibnamefont
  {Schmidt}},\ }\bibfield  {title} {\enquote {\bibinfo {title} {Phase diagram
  for strong-coupling bose polarons},}\ }\href@noop {} {\bibfield  {journal}
  {\bibinfo  {journal} {arXiv preprint arXiv:2306.09075}\ } (\bibinfo {year}
  {2023})}\BibitemShut {NoStop}%
\bibitem [{\citenamefont {Levinsen}\ \emph {et~al.}(2021)\citenamefont
  {Levinsen}, \citenamefont {Ardila}, \citenamefont {Yoshida},\ and\
  \citenamefont {Parish}}]{levinsen2021quantum}%
  \BibitemOpen
  \bibfield  {author} {\bibinfo {author} {\bibfnamefont {Jesper}\ \bibnamefont
  {Levinsen}}, \bibinfo {author} {\bibfnamefont {Luis A~Pe{\~n}a}\ \bibnamefont
  {Ardila}}, \bibinfo {author} {\bibfnamefont {Shuhei~M}\ \bibnamefont
  {Yoshida}}, \ and\ \bibinfo {author} {\bibfnamefont {Meera~M}\ \bibnamefont
  {Parish}},\ }\bibfield  {title} {\enquote {\bibinfo {title} {Quantum behavior
  of a heavy impurity strongly coupled to a bose gas},}\ }\href@noop {}
  {\bibfield  {journal} {\bibinfo  {journal} {Physical Review Letters}\
  }\textbf {\bibinfo {volume} {127}},\ \bibinfo {pages} {033401} (\bibinfo
  {year} {2021})}\BibitemShut {NoStop}%
\bibitem [{\citenamefont {Levinsen}\ \emph {et~al.}(2017)\citenamefont
  {Levinsen}, \citenamefont {Parish}, \citenamefont {Christensen},
  \citenamefont {Arlt},\ and\ \citenamefont {Bruun}}]{Levinsen2017}%
  \BibitemOpen
  \bibfield  {author} {\bibinfo {author} {\bibfnamefont {Jesper}\ \bibnamefont
  {Levinsen}}, \bibinfo {author} {\bibfnamefont {Meera~M.}\ \bibnamefont
  {Parish}}, \bibinfo {author} {\bibfnamefont {Rasmus~S.}\ \bibnamefont
  {Christensen}}, \bibinfo {author} {\bibfnamefont {Jan~J.}\ \bibnamefont
  {Arlt}}, \ and\ \bibinfo {author} {\bibfnamefont {Georg~M.}\ \bibnamefont
  {Bruun}},\ }\bibfield  {title} {\enquote {\bibinfo {title}
  {Finite-temperature behavior of the bose polaron},}\ }\href {\doibase
  10.1103/PhysRevA.96.063622} {\bibfield  {journal} {\bibinfo  {journal} {Phys.
  Rev. A}\ }\textbf {\bibinfo {volume} {96}},\ \bibinfo {pages} {063622}
  (\bibinfo {year} {2017})}\BibitemShut {NoStop}%
\bibitem [{\citenamefont {Guenther}\ \emph {et~al.}(2018)\citenamefont
  {Guenther}, \citenamefont {Massignan}, \citenamefont {Lewenstein},\ and\
  \citenamefont {Bruun}}]{guenther2018bose}%
  \BibitemOpen
  \bibfield  {author} {\bibinfo {author} {\bibfnamefont {Nils-Eric}\
  \bibnamefont {Guenther}}, \bibinfo {author} {\bibfnamefont {Pietro}\
  \bibnamefont {Massignan}}, \bibinfo {author} {\bibfnamefont {Maciej}\
  \bibnamefont {Lewenstein}}, \ and\ \bibinfo {author} {\bibfnamefont
  {Georg~M}\ \bibnamefont {Bruun}},\ }\bibfield  {title} {\enquote {\bibinfo
  {title} {Bose polarons at finite temperature and strong coupling},}\
  }\href@noop {} {\bibfield  {journal} {\bibinfo  {journal} {Physical review
  letters}\ }\textbf {\bibinfo {volume} {120}},\ \bibinfo {pages} {050405}
  (\bibinfo {year} {2018})}\BibitemShut {NoStop}%
\bibitem [{\citenamefont {Field}\ \emph {et~al.}(2020)\citenamefont {Field},
  \citenamefont {Levinsen},\ and\ \citenamefont {Parish}}]{field2020fate}%
  \BibitemOpen
  \bibfield  {author} {\bibinfo {author} {\bibfnamefont {Bernard}\ \bibnamefont
  {Field}}, \bibinfo {author} {\bibfnamefont {Jesper}\ \bibnamefont
  {Levinsen}}, \ and\ \bibinfo {author} {\bibfnamefont {Meera~M}\ \bibnamefont
  {Parish}},\ }\bibfield  {title} {\enquote {\bibinfo {title} {Fate of the bose
  polaron at finite temperature},}\ }\href@noop {} {\bibfield  {journal}
  {\bibinfo  {journal} {Physical Review A}\ }\textbf {\bibinfo {volume}
  {101}},\ \bibinfo {pages} {013623} (\bibinfo {year} {2020})}\BibitemShut
  {NoStop}%
\bibitem [{\citenamefont {Drescher}\ \emph {et~al.}(2021)\citenamefont
  {Drescher}, \citenamefont {Salmhofer},\ and\ \citenamefont
  {Enss}}]{Drescher2021}%
  \BibitemOpen
  \bibfield  {author} {\bibinfo {author} {\bibfnamefont {Moritz}\ \bibnamefont
  {Drescher}}, \bibinfo {author} {\bibfnamefont {Manfred}\ \bibnamefont
  {Salmhofer}}, \ and\ \bibinfo {author} {\bibfnamefont {Tilman}\ \bibnamefont
  {Enss}},\ }\bibfield  {title} {\enquote {\bibinfo {title} {Quench dynamics of
  the ideal bose polaron at zero and nonzero temperatures},}\ }\href {\doibase
  10.1103/PhysRevA.103.033317} {\bibfield  {journal} {\bibinfo  {journal}
  {Phys. Rev. A}\ }\textbf {\bibinfo {volume} {103}},\ \bibinfo {pages}
  {033317} (\bibinfo {year} {2021})}\BibitemShut {NoStop}%
\bibitem [{\citenamefont {Hryhorchak}\ and\ \citenamefont
  {Pastukhov}(2023)}]{hryhorchak2023trapped}%
  \BibitemOpen
  \bibfield  {author} {\bibinfo {author} {\bibfnamefont {Orest}\ \bibnamefont
  {Hryhorchak}}\ and\ \bibinfo {author} {\bibfnamefont {Volodymyr}\
  \bibnamefont {Pastukhov}},\ }\bibfield  {title} {\enquote {\bibinfo {title}
  {Trapped ideal bose gas with a few heavy impurities},}\ }\href@noop {}
  {\bibfield  {journal} {\bibinfo  {journal} {Atoms}\ }\textbf {\bibinfo
  {volume} {11}},\ \bibinfo {pages} {77} (\bibinfo {year} {2023})}\BibitemShut
  {NoStop}%
\bibitem [{\citenamefont {Isaule}(2024)}]{isaule2024functional}%
  \BibitemOpen
  \bibfield  {author} {\bibinfo {author} {\bibfnamefont {Felipe}\ \bibnamefont
  {Isaule}},\ }\bibfield  {title} {\enquote {\bibinfo {title} {Functional
  renormalisation group approach to the finite-temperature bose polaron},}\
  }\href@noop {} {\bibfield  {journal} {\bibinfo  {journal} {arXiv preprint
  arXiv:2402.04197}\ } (\bibinfo {year} {2024})}\BibitemShut {NoStop}%
\bibitem [{\citenamefont {Ardila}\ and\ \citenamefont
  {Pohl}(2018)}]{Ardila2018}%
  \BibitemOpen
  \bibfield  {author} {\bibinfo {author} {\bibfnamefont {L~A~Pe{\~{n}}a}\
  \bibnamefont {Ardila}}\ and\ \bibinfo {author} {\bibfnamefont
  {T}~\bibnamefont {Pohl}},\ }\bibfield  {title} {\enquote {\bibinfo {title}
  {Ground-state properties of dipolar bose polarons},}\ }\href {\doibase
  10.1088/1361-6455/aaf35e} {\bibfield  {journal} {\bibinfo  {journal} {Journal
  of Physics B: Atomic, Molecular and Optical Physics}\ }\textbf {\bibinfo
  {volume} {52}},\ \bibinfo {pages} {015004} (\bibinfo {year}
  {2018})}\BibitemShut {NoStop}%
\bibitem [{\citenamefont {Ardila}(2022)}]{Ardila2022N}%
  \BibitemOpen
  \bibfield  {author} {\bibinfo {author} {\bibfnamefont {L.~A.~P.}\
  \bibnamefont {Ardila}},\ }\bibfield  {title} {\enquote {\bibinfo {title}
  {Monte carlo methods for impurity physics in ultracold bose quantum gases},}\
  }\href {\doibase 10.1038/s42254-022-00443-5} {\bibfield  {journal} {\bibinfo
  {journal} {Nature Reviews Physics}\ }\textbf {\bibinfo {volume} {4}},\
  \bibinfo {pages} {214--214} (\bibinfo {year} {2022})}\BibitemShut {NoStop}%
\bibitem [{\citenamefont {Astrakharchik}\ \emph {et~al.}(2021)\citenamefont
  {Astrakharchik}, \citenamefont {Ardila}, \citenamefont {Schmidt},
  \citenamefont {Jachymski},\ and\ \citenamefont
  {Negretti}}]{Astrakharchik2021}%
  \BibitemOpen
  \bibfield  {author} {\bibinfo {author} {\bibfnamefont {Grigory~E}\
  \bibnamefont {Astrakharchik}}, \bibinfo {author} {\bibfnamefont {Luis
  A~Pe{\~n}a}\ \bibnamefont {Ardila}}, \bibinfo {author} {\bibfnamefont
  {Richard}\ \bibnamefont {Schmidt}}, \bibinfo {author} {\bibfnamefont
  {Krzysztof}\ \bibnamefont {Jachymski}}, \ and\ \bibinfo {author}
  {\bibfnamefont {Antonio}\ \bibnamefont {Negretti}},\ }\bibfield  {title}
  {\enquote {\bibinfo {title} {Ionic polaron in a bose-einstein condensate},}\
  }\href@noop {} {\bibfield  {journal} {\bibinfo  {journal} {Communications
  Physics}\ }\textbf {\bibinfo {volume} {4}},\ \bibinfo {pages} {1--8}
  (\bibinfo {year} {2021})}\BibitemShut {NoStop}%
\bibitem [{\citenamefont {Christensen}\ \emph {et~al.}(2021)\citenamefont
  {Christensen}, \citenamefont {Camacho-Guardian},\ and\ \citenamefont
  {Bruun}}]{Christensen2021}%
  \BibitemOpen
  \bibfield  {author} {\bibinfo {author} {\bibfnamefont {Esben~Rohan}\
  \bibnamefont {Christensen}}, \bibinfo {author} {\bibfnamefont {Arturo}\
  \bibnamefont {Camacho-Guardian}}, \ and\ \bibinfo {author} {\bibfnamefont
  {Georg~M.}\ \bibnamefont {Bruun}},\ }\bibfield  {title} {\enquote {\bibinfo
  {title} {Charged polarons and molecules in a bose-einstein condensate},}\
  }\href {\doibase 10.1103/PhysRevLett.126.243001} {\bibfield  {journal}
  {\bibinfo  {journal} {Phys. Rev. Lett.}\ }\textbf {\bibinfo {volume} {126}},\
  \bibinfo {pages} {243001} (\bibinfo {year} {2021})}\BibitemShut {NoStop}%
\bibitem [{\citenamefont {{Ding}}\ \emph {et~al.}(2022)\citenamefont {{Ding}},
  \citenamefont {{Drewsen}}, \citenamefont {{Arlt}},\ and\ \citenamefont
  {{Bruun}}}]{Ding2022}%
  \BibitemOpen
  \bibfield  {author} {\bibinfo {author} {\bibfnamefont {Shanshan}\
  \bibnamefont {{Ding}}}, \bibinfo {author} {\bibfnamefont {Michael}\
  \bibnamefont {{Drewsen}}}, \bibinfo {author} {\bibfnamefont {Jan~J.}\
  \bibnamefont {{Arlt}}}, \ and\ \bibinfo {author} {\bibfnamefont {G.~M.}\
  \bibnamefont {{Bruun}}},\ }\bibfield  {title} {\enquote {\bibinfo {title}
  {{Mediated interactions between ions in quantum degenerate gases}},}\
  }\href@noop {} {\bibfield  {journal} {\bibinfo  {journal} {arXiv e-prints}\
  ,\ \bibinfo {eid} {arXiv:2203.02768}} (\bibinfo {year} {2022})},\ \Eprint
  {http://arxiv.org/abs/2203.02768} {arXiv:2203.02768 [cond-mat.quant-gas]}
  \BibitemShut {NoStop}%
\bibitem [{\citenamefont {{Astrakharchik}}\ \emph {et~al.}(2022)\citenamefont
  {{Astrakharchik}}, \citenamefont {{Pe{\~n}a Ardila}}, \citenamefont
  {{Jachymski}},\ and\ \citenamefont {{Negretti}}}]{Astrakharchik2022}%
  \BibitemOpen
  \bibfield  {author} {\bibinfo {author} {\bibfnamefont {G.~E.}\ \bibnamefont
  {{Astrakharchik}}}, \bibinfo {author} {\bibfnamefont {L.~A.}\ \bibnamefont
  {{Pe{\~n}a Ardila}}}, \bibinfo {author} {\bibfnamefont {K.}~\bibnamefont
  {{Jachymski}}}, \ and\ \bibinfo {author} {\bibfnamefont {A.}~\bibnamefont
  {{Negretti}}},\ }\bibfield  {title} {\enquote {\bibinfo {title} {{Charged
  impurities in a Bose-Einstein condensate: Many-body bound states and induced
  interactions}},}\ }\href@noop {} {\bibfield  {journal} {\bibinfo  {journal}
  {arXiv e-prints}\ ,\ \bibinfo {eid} {arXiv:2206.03476}} (\bibinfo {year}
  {2022})},\ \Eprint {http://arxiv.org/abs/2206.03476} {arXiv:2206.03476
  [cond-mat.quant-gas]} \BibitemShut {NoStop}%
\bibitem [{\citenamefont {Cavazos~Olivas}\ \emph {et~al.}(2024)\citenamefont
  {Cavazos~Olivas}, \citenamefont {Pe\~na Ardila},\ and\ \citenamefont
  {Jachymski}}]{olivas2024}%
  \BibitemOpen
  \bibfield  {author} {\bibinfo {author} {\bibfnamefont {Ubaldo}\ \bibnamefont
  {Cavazos~Olivas}}, \bibinfo {author} {\bibfnamefont {Luis~A.}\ \bibnamefont
  {Pe\~na Ardila}}, \ and\ \bibinfo {author} {\bibfnamefont {Krzysztof}\
  \bibnamefont {Jachymski}},\ }\bibfield  {title} {\enquote {\bibinfo {title}
  {Modified mean-field ansatz for charged polarons in a bose-einstein
  condensate},}\ }\href {\doibase 10.1103/PhysRevA.110.L011301} {\bibfield
  {journal} {\bibinfo  {journal} {Phys. Rev. A}\ }\textbf {\bibinfo {volume}
  {110}},\ \bibinfo {pages} {L011301} (\bibinfo {year} {2024})}\BibitemShut
  {NoStop}%
\bibitem [{\citenamefont {Volosniev}\ \emph {et~al.}(2023)\citenamefont
  {Volosniev}, \citenamefont {Bighin}, \citenamefont {Santos},\ and\
  \citenamefont {Ardila}}]{volosniev2023non}%
  \BibitemOpen
  \bibfield  {author} {\bibinfo {author} {\bibfnamefont {Artem~G}\ \bibnamefont
  {Volosniev}}, \bibinfo {author} {\bibfnamefont {Giacomo}\ \bibnamefont
  {Bighin}}, \bibinfo {author} {\bibfnamefont {Luis}\ \bibnamefont {Santos}}, \
  and\ \bibinfo {author} {\bibfnamefont {Luis A~Pe{\~n}a}\ \bibnamefont
  {Ardila}},\ }\bibfield  {title} {\enquote {\bibinfo {title} {Non-equilibrium
  dynamics of dipolar polarons},}\ }\href@noop {} {\bibfield  {journal}
  {\bibinfo  {journal} {arXiv preprint arXiv:2305.17969}\ } (\bibinfo {year}
  {2023})}\BibitemShut {NoStop}%
\bibitem [{\citenamefont {Levinsen}\ \emph {et~al.}(2015)\citenamefont
  {Levinsen}, \citenamefont {Parish},\ and\ \citenamefont
  {Bruun}}]{levinsen2015impurity}%
  \BibitemOpen
  \bibfield  {author} {\bibinfo {author} {\bibfnamefont {Jesper}\ \bibnamefont
  {Levinsen}}, \bibinfo {author} {\bibfnamefont {Meera~M}\ \bibnamefont
  {Parish}}, \ and\ \bibinfo {author} {\bibfnamefont {Georg~M}\ \bibnamefont
  {Bruun}},\ }\bibfield  {title} {\enquote {\bibinfo {title} {Impurity in a
  bose-einstein condensate and the efimov effect},}\ }\href@noop {} {\bibfield
  {journal} {\bibinfo  {journal} {Physical Review Letters}\ }\textbf {\bibinfo
  {volume} {115}},\ \bibinfo {pages} {125302} (\bibinfo {year}
  {2015})}\BibitemShut {NoStop}%
\bibitem [{\citenamefont {Sun}\ \emph {et~al.}(2017)\citenamefont {Sun},
  \citenamefont {Zhai},\ and\ \citenamefont {Cui}}]{Sun2017}%
  \BibitemOpen
  \bibfield  {author} {\bibinfo {author} {\bibfnamefont {Mingyuan}\
  \bibnamefont {Sun}}, \bibinfo {author} {\bibfnamefont {Hui}\ \bibnamefont
  {Zhai}}, \ and\ \bibinfo {author} {\bibfnamefont {Xiaoling}\ \bibnamefont
  {Cui}},\ }\bibfield  {title} {\enquote {\bibinfo {title} {Visualizing the
  efimov correlation in bose polarons},}\ }\href {\doibase
  10.1103/PhysRevLett.119.013401} {\bibfield  {journal} {\bibinfo  {journal}
  {Phys. Rev. Lett.}\ }\textbf {\bibinfo {volume} {119}},\ \bibinfo {pages}
  {013401} (\bibinfo {year} {2017})}\BibitemShut {NoStop}%
\bibitem [{\citenamefont {Sun}\ and\ \citenamefont {Cui}(2017)}]{Sun2017b}%
  \BibitemOpen
  \bibfield  {author} {\bibinfo {author} {\bibfnamefont {Mingyuan}\
  \bibnamefont {Sun}}\ and\ \bibinfo {author} {\bibfnamefont {Xiaoling}\
  \bibnamefont {Cui}},\ }\bibfield  {title} {\enquote {\bibinfo {title}
  {Enhancing the efimov correlation in bose polarons with large mass
  imbalance},}\ }\href {\doibase 10.1103/PhysRevA.96.022707} {\bibfield
  {journal} {\bibinfo  {journal} {Phys. Rev. A}\ }\textbf {\bibinfo {volume}
  {96}},\ \bibinfo {pages} {022707} (\bibinfo {year} {2017})}\BibitemShut
  {NoStop}%
\bibitem [{\citenamefont {Christianen}\ \emph
  {et~al.}(2022{\natexlab{a}})\citenamefont {Christianen}, \citenamefont
  {Cirac},\ and\ \citenamefont {Schmidt}}]{christianen2022chemistry}%
  \BibitemOpen
  \bibfield  {author} {\bibinfo {author} {\bibfnamefont {Arthur}\ \bibnamefont
  {Christianen}}, \bibinfo {author} {\bibfnamefont {J~Ignacio}\ \bibnamefont
  {Cirac}}, \ and\ \bibinfo {author} {\bibfnamefont {Richard}\ \bibnamefont
  {Schmidt}},\ }\bibfield  {title} {\enquote {\bibinfo {title} {Chemistry of a
  light impurity in a bose-einstein condensate},}\ }\href@noop {} {\bibfield
  {journal} {\bibinfo  {journal} {Physical Review Letters}\ }\textbf {\bibinfo
  {volume} {128}},\ \bibinfo {pages} {183401} (\bibinfo {year}
  {2022}{\natexlab{a}})}\BibitemShut {NoStop}%
\bibitem [{\citenamefont {Christianen}\ \emph
  {et~al.}(2022{\natexlab{b}})\citenamefont {Christianen}, \citenamefont
  {Cirac},\ and\ \citenamefont {Schmidt}}]{christianen2022bose}%
  \BibitemOpen
  \bibfield  {author} {\bibinfo {author} {\bibfnamefont {Arthur}\ \bibnamefont
  {Christianen}}, \bibinfo {author} {\bibfnamefont {J~Ignacio}\ \bibnamefont
  {Cirac}}, \ and\ \bibinfo {author} {\bibfnamefont {Richard}\ \bibnamefont
  {Schmidt}},\ }\bibfield  {title} {\enquote {\bibinfo {title} {Bose polaron
  and the efimov effect: A gaussian-state approach},}\ }\href@noop {}
  {\bibfield  {journal} {\bibinfo  {journal} {Physical Review A}\ }\textbf
  {\bibinfo {volume} {105}},\ \bibinfo {pages} {053302} (\bibinfo {year}
  {2022}{\natexlab{b}})}\BibitemShut {NoStop}%
\bibitem [{\citenamefont {Yoshida}\ \emph {et~al.}(2018)\citenamefont
  {Yoshida}, \citenamefont {Endo}, \citenamefont {Levinsen},\ and\
  \citenamefont {Parish}}]{Yoshida2018}%
  \BibitemOpen
  \bibfield  {author} {\bibinfo {author} {\bibfnamefont {Shuhei~M.}\
  \bibnamefont {Yoshida}}, \bibinfo {author} {\bibfnamefont {Shimpei}\
  \bibnamefont {Endo}}, \bibinfo {author} {\bibfnamefont {Jesper}\ \bibnamefont
  {Levinsen}}, \ and\ \bibinfo {author} {\bibfnamefont {Meera~M.}\ \bibnamefont
  {Parish}},\ }\bibfield  {title} {\enquote {\bibinfo {title} {Universality of
  an impurity in a bose-einstein condensate},}\ }\href {\doibase
  10.1103/PhysRevX.8.011024} {\bibfield  {journal} {\bibinfo  {journal} {Phys.
  Rev. X}\ }\textbf {\bibinfo {volume} {8}},\ \bibinfo {pages} {011024}
  (\bibinfo {year} {2018})}\BibitemShut {NoStop}%
\bibitem [{\citenamefont {Massignan}\ \emph {et~al.}(2021)\citenamefont
  {Massignan}, \citenamefont {Yegovtsev},\ and\ \citenamefont
  {Gurarie}}]{Massignan2021}%
  \BibitemOpen
  \bibfield  {author} {\bibinfo {author} {\bibfnamefont {Pietro}\ \bibnamefont
  {Massignan}}, \bibinfo {author} {\bibfnamefont {Nikolay}\ \bibnamefont
  {Yegovtsev}}, \ and\ \bibinfo {author} {\bibfnamefont {Victor}\ \bibnamefont
  {Gurarie}},\ }\bibfield  {title} {\enquote {\bibinfo {title} {Universal
  aspects of a strongly interacting impurity in a dilute bose condensate},}\
  }\href {\doibase 10.1103/PhysRevLett.126.123403} {\bibfield  {journal}
  {\bibinfo  {journal} {Phys. Rev. Lett.}\ }\textbf {\bibinfo {volume} {126}},\
  \bibinfo {pages} {123403} (\bibinfo {year} {2021})}\BibitemShut {NoStop}%
\bibitem [{\citenamefont {Yegovtsev}\ \emph {et~al.}(2022)\citenamefont
  {Yegovtsev}, \citenamefont {Massignan},\ and\ \citenamefont
  {Gurarie}}]{Yegovtsev2022}%
  \BibitemOpen
  \bibfield  {author} {\bibinfo {author} {\bibfnamefont {Nikolay}\ \bibnamefont
  {Yegovtsev}}, \bibinfo {author} {\bibfnamefont {Pietro}\ \bibnamefont
  {Massignan}}, \ and\ \bibinfo {author} {\bibfnamefont {Victor}\ \bibnamefont
  {Gurarie}},\ }\bibfield  {title} {\enquote {\bibinfo {title} {Strongly
  interacting impurities in a dilute bose condensate},}\ }\href {\doibase
  10.1103/PhysRevA.106.033305} {\bibfield  {journal} {\bibinfo  {journal}
  {Phys. Rev. A}\ }\textbf {\bibinfo {volume} {106}},\ \bibinfo {pages}
  {033305} (\bibinfo {year} {2022})}\BibitemShut {NoStop}%
\bibitem [{\citenamefont {Skou}\ \emph
  {et~al.}(2021{\natexlab{b}})\citenamefont {Skou}, \citenamefont {Skov},
  \citenamefont {J{\o}rgensen},\ and\ \citenamefont {Arlt}}]{atoms9020022}%
  \BibitemOpen
  \bibfield  {author} {\bibinfo {author} {\bibfnamefont {Magnus~G.}\
  \bibnamefont {Skou}}, \bibinfo {author} {\bibfnamefont {Thomas~G.}\
  \bibnamefont {Skov}}, \bibinfo {author} {\bibfnamefont {Nils~B.}\
  \bibnamefont {J{\o}rgensen}}, \ and\ \bibinfo {author} {\bibfnamefont
  {Jan~J.}\ \bibnamefont {Arlt}},\ }\bibfield  {title} {\enquote {\bibinfo
  {title} {Initial dynamics of quantum impurities in a bose--einstein
  condensate},}\ }\href {\doibase 10.3390/atoms9020022} {\bibfield  {journal}
  {\bibinfo  {journal} {Atoms}\ }\textbf {\bibinfo {volume} {9}} (\bibinfo
  {year} {2021}{\natexlab{b}}),\ 10.3390/atoms9020022}\BibitemShut {NoStop}%
\bibitem [{\citenamefont {Cayla}\ \emph {et~al.}(2023)\citenamefont {Cayla},
  \citenamefont {Massignan}, \citenamefont {Giamarchi}, \citenamefont {Aspect},
  \citenamefont {Westbrook},\ and\ \citenamefont
  {Cl{\'e}ment}}]{cayla2023observation}%
  \BibitemOpen
  \bibfield  {author} {\bibinfo {author} {\bibfnamefont {Hugo}\ \bibnamefont
  {Cayla}}, \bibinfo {author} {\bibfnamefont {Pietro}\ \bibnamefont
  {Massignan}}, \bibinfo {author} {\bibfnamefont {Thierry}\ \bibnamefont
  {Giamarchi}}, \bibinfo {author} {\bibfnamefont {Alain}\ \bibnamefont
  {Aspect}}, \bibinfo {author} {\bibfnamefont {Christoph~I}\ \bibnamefont
  {Westbrook}}, \ and\ \bibinfo {author} {\bibfnamefont {David}\ \bibnamefont
  {Cl{\'e}ment}},\ }\bibfield  {title} {\enquote {\bibinfo {title} {Observation
  of 1/k 4-tails after expansion of bose-einstein condensates with
  impurities},}\ }\href@noop {} {\bibfield  {journal} {\bibinfo  {journal}
  {Physical Review Letters}\ }\textbf {\bibinfo {volume} {130}},\ \bibinfo
  {pages} {153401} (\bibinfo {year} {2023})}\BibitemShut {NoStop}%
\bibitem [{\citenamefont {Colussi}\ \emph {et~al.}(2023)\citenamefont
  {Colussi}, \citenamefont {Caleffi}, \citenamefont {Menotti},\ and\
  \citenamefont {Recati}}]{colussi2023lattice}%
  \BibitemOpen
  \bibfield  {author} {\bibinfo {author} {\bibfnamefont {VE}~\bibnamefont
  {Colussi}}, \bibinfo {author} {\bibfnamefont {F}~\bibnamefont {Caleffi}},
  \bibinfo {author} {\bibfnamefont {C}~\bibnamefont {Menotti}}, \ and\ \bibinfo
  {author} {\bibfnamefont {A}~\bibnamefont {Recati}},\ }\bibfield  {title}
  {\enquote {\bibinfo {title} {Lattice polarons across the superfluid to mott
  insulator transition},}\ }\href@noop {} {\bibfield  {journal} {\bibinfo
  {journal} {Physical Review Letters}\ }\textbf {\bibinfo {volume} {130}},\
  \bibinfo {pages} {173002} (\bibinfo {year} {2023})}\BibitemShut {NoStop}%
\bibitem [{\citenamefont {Ding}\ \emph {et~al.}(2023)\citenamefont {Ding},
  \citenamefont {Dom{\'\i}nguez-Castro}, \citenamefont {Julku}, \citenamefont
  {Camacho~Guardian},\ and\ \citenamefont {Bruun}}]{ding2023polarons}%
  \BibitemOpen
  \bibfield  {author} {\bibinfo {author} {\bibfnamefont {Shanshan}\
  \bibnamefont {Ding}}, \bibinfo {author} {\bibfnamefont {GA}~\bibnamefont
  {Dom{\'\i}nguez-Castro}}, \bibinfo {author} {\bibfnamefont {Aleksi}\
  \bibnamefont {Julku}}, \bibinfo {author} {\bibfnamefont {Arturo}\
  \bibnamefont {Camacho~Guardian}}, \ and\ \bibinfo {author} {\bibfnamefont
  {Georg~M}\ \bibnamefont {Bruun}},\ }\bibfield  {title} {\enquote {\bibinfo
  {title} {Polarons and bipolarons in a two-dimensional square lattice},}\
  }\href@noop {} {\bibfield  {journal} {\bibinfo  {journal} {SciPost Physics}\
  }\textbf {\bibinfo {volume} {14}},\ \bibinfo {pages} {143} (\bibinfo {year}
  {2023})}\BibitemShut {NoStop}%
\bibitem [{\citenamefont {Isaule}\ \emph {et~al.}(2024)\citenamefont {Isaule},
  \citenamefont {Rojo-Franc{\`a}s},\ and\ \citenamefont
  {Juli{\'a}-D{\'\i}az}}]{isaule2024bound}%
  \BibitemOpen
  \bibfield  {author} {\bibinfo {author} {\bibfnamefont {Felipe}\ \bibnamefont
  {Isaule}}, \bibinfo {author} {\bibfnamefont {Abel}\ \bibnamefont
  {Rojo-Franc{\`a}s}}, \ and\ \bibinfo {author} {\bibfnamefont {Bruno}\
  \bibnamefont {Juli{\'a}-D{\'\i}az}},\ }\bibfield  {title} {\enquote {\bibinfo
  {title} {Bound impurities in a one-dimensional bose lattice gas: low-energy
  properties and quench-induced dynamics},}\ }\href@noop {} {\bibfield
  {journal} {\bibinfo  {journal} {arXiv preprint arXiv:2402.03070}\ } (\bibinfo
  {year} {2024})}\BibitemShut {NoStop}%
\bibitem [{\citenamefont {Amelio}\ \emph {et~al.}(2024)\citenamefont {Amelio},
  \citenamefont {Mazza},\ and\ \citenamefont {Goldman}}]{amelio2024polaron}%
  \BibitemOpen
  \bibfield  {author} {\bibinfo {author} {\bibfnamefont {Ivan}\ \bibnamefont
  {Amelio}}, \bibinfo {author} {\bibfnamefont {Giacomo}\ \bibnamefont {Mazza}},
  \ and\ \bibinfo {author} {\bibfnamefont {Nathan}\ \bibnamefont {Goldman}},\
  }\bibfield  {title} {\enquote {\bibinfo {title} {Polaron formation in
  insulators and the key role of hole scattering processes: Band insulators,
  charge density waves and mott transition},}\ }\href@noop {} {\bibfield
  {journal} {\bibinfo  {journal} {arXiv preprint arXiv:2408.01377}\ } (\bibinfo
  {year} {2024})}\BibitemShut {NoStop}%
\bibitem [{\citenamefont {Santiago-García}\ and\ \citenamefont
  {Camacho-Guardian}(2023)}]{Santiago-Garcia_2023}%
  \BibitemOpen
  \bibfield  {author} {\bibinfo {author} {\bibfnamefont {Moroni}\ \bibnamefont
  {Santiago-García}}\ and\ \bibinfo {author} {\bibfnamefont {Arturo}\
  \bibnamefont {Camacho-Guardian}},\ }\bibfield  {title} {\enquote {\bibinfo
  {title} {Collective excitations of a bose–einstein condensate of hard-core
  bosons and their mediated interactions: from two-body bound states to
  mediated superfluidity},}\ }\href {\doibase 10.1088/1367-2630/acf72d}
  {\bibfield  {journal} {\bibinfo  {journal} {New Journal of Physics}\ }\textbf
  {\bibinfo {volume} {25}},\ \bibinfo {pages} {093032} (\bibinfo {year}
  {2023})}\BibitemShut {NoStop}%
\bibitem [{\citenamefont {Santiago-García}\ \emph {et~al.}(2024)\citenamefont
  {Santiago-García}, \citenamefont {Castillo-López},\ and\ \citenamefont
  {Camacho-Guardian}}]{santiago2024lattice}%
  \BibitemOpen
  \bibfield  {author} {\bibinfo {author} {\bibfnamefont {Moroni}\ \bibnamefont
  {Santiago-García}}, \bibinfo {author} {\bibfnamefont {Shunashi~G}\
  \bibnamefont {Castillo-López}}, \ and\ \bibinfo {author} {\bibfnamefont
  {Arturo}\ \bibnamefont {Camacho-Guardian}},\ }\bibfield  {title} {\enquote
  {\bibinfo {title} {Lattice polaron in a bose–einstein condensate of
  hard-core bosons},}\ }\href {\doibase 10.1088/1367-2630/ad503e} {\bibfield
  {journal} {\bibinfo  {journal} {New Journal of Physics}\ }\textbf {\bibinfo
  {volume} {26}},\ \bibinfo {pages} {063015} (\bibinfo {year}
  {2024})}\BibitemShut {NoStop}%
\bibitem [{\citenamefont {Pimenov}(2024)}]{Dima2024}%
  \BibitemOpen
  \bibfield  {author} {\bibinfo {author} {\bibfnamefont {Dimitri}\ \bibnamefont
  {Pimenov}},\ }\bibfield  {title} {\enquote {\bibinfo {title} {Polaron spectra
  and edge singularities for correlated flat bands},}\ }\href {\doibase
  10.1103/PhysRevB.109.195153} {\bibfield  {journal} {\bibinfo  {journal}
  {Phys. Rev. B}\ }\textbf {\bibinfo {volume} {109}},\ \bibinfo {pages}
  {195153} (\bibinfo {year} {2024})}\BibitemShut {NoStop}%
\bibitem [{\citenamefont {Vashisht}\ \emph {et~al.}(2024)\citenamefont
  {Vashisht}, \citenamefont {Amelio}, \citenamefont {Vanderstraeten},
  \citenamefont {Bruun}, \citenamefont {Diessel},\ and\ \citenamefont
  {Goldman}}]{vashisht2024chiral}%
  \BibitemOpen
  \bibfield  {author} {\bibinfo {author} {\bibfnamefont {Amit}\ \bibnamefont
  {Vashisht}}, \bibinfo {author} {\bibfnamefont {Ivan}\ \bibnamefont {Amelio}},
  \bibinfo {author} {\bibfnamefont {Laurens}\ \bibnamefont {Vanderstraeten}},
  \bibinfo {author} {\bibfnamefont {Georg~M}\ \bibnamefont {Bruun}}, \bibinfo
  {author} {\bibfnamefont {Oriana~K}\ \bibnamefont {Diessel}}, \ and\ \bibinfo
  {author} {\bibfnamefont {Nathan}\ \bibnamefont {Goldman}},\ }\bibfield
  {title} {\enquote {\bibinfo {title} {Chiral polaron formation on the edge of
  topological quantum matter},}\ }\href@noop {} {\bibfield  {journal} {\bibinfo
   {journal} {arXiv preprint arXiv:2407.19093}\ } (\bibinfo {year}
  {2024})}\BibitemShut {NoStop}%
\bibitem [{\citenamefont {Pimenov}\ \emph {et~al.}(2021)\citenamefont
  {Pimenov}, \citenamefont {Camacho-Guardian}, \citenamefont {Goldman},
  \citenamefont {Massignan}, \citenamefont {Bruun},\ and\ \citenamefont
  {Goldstein}}]{Dima2021}%
  \BibitemOpen
  \bibfield  {author} {\bibinfo {author} {\bibfnamefont {D.}~\bibnamefont
  {Pimenov}}, \bibinfo {author} {\bibfnamefont {A.}~\bibnamefont
  {Camacho-Guardian}}, \bibinfo {author} {\bibfnamefont {N.}~\bibnamefont
  {Goldman}}, \bibinfo {author} {\bibfnamefont {P.}~\bibnamefont {Massignan}},
  \bibinfo {author} {\bibfnamefont {G.~M.}\ \bibnamefont {Bruun}}, \ and\
  \bibinfo {author} {\bibfnamefont {M.}~\bibnamefont {Goldstein}},\ }\bibfield
  {title} {\enquote {\bibinfo {title} {Topological transport of mobile
  impurities},}\ }\href {\doibase 10.1103/PhysRevB.103.245106} {\bibfield
  {journal} {\bibinfo  {journal} {Phys. Rev. B}\ }\textbf {\bibinfo {volume}
  {103}},\ \bibinfo {pages} {245106} (\bibinfo {year} {2021})}\BibitemShut
  {NoStop}%
\bibitem [{\citenamefont {{Christ}}\ \emph {et~al.}(2024)\citenamefont
  {{Christ}}, \citenamefont {{Bermes}},\ and\ \citenamefont
  {{Grusdt}}}]{Christ2024}%
  \BibitemOpen
  \bibfield  {author} {\bibinfo {author} {\bibfnamefont {Jan-Philipp}\
  \bibnamefont {{Christ}}}, \bibinfo {author} {\bibfnamefont {Pit}\
  \bibnamefont {{Bermes}}}, \ and\ \bibinfo {author} {\bibfnamefont {Fabian}\
  \bibnamefont {{Grusdt}}},\ }\bibfield  {title} {\enquote {\bibinfo {title}
  {{Operator Valued Flow Equation Approach to the Bosonic Lattice Polaron:
  Dispersion Renormalization Beyond the Fr{\"o}hlich Paradigm}},}\ }\href
  {\doibase 10.48550/arXiv.2411.19947} {\bibfield  {journal} {\bibinfo
  {journal} {arXiv e-prints}\ ,\ \bibinfo {eid} {arXiv:2411.19947}} (\bibinfo
  {year} {2024})},\ \Eprint {http://arxiv.org/abs/2411.19947} {arXiv:2411.19947
  [cond-mat.quant-gas]} \BibitemShut {NoStop}%
\bibitem [{\citenamefont {Schmidt}\ and\ \citenamefont
  {Lemeshko}(2015)}]{Schmidt2015}%
  \BibitemOpen
  \bibfield  {author} {\bibinfo {author} {\bibfnamefont {Richard}\ \bibnamefont
  {Schmidt}}\ and\ \bibinfo {author} {\bibfnamefont {Mikhail}\ \bibnamefont
  {Lemeshko}},\ }\bibfield  {title} {\enquote {\bibinfo {title} {Rotation of
  quantum impurities in the presence of a many-body environment},}\ }\href
  {\doibase 10.1103/PhysRevLett.114.203001} {\bibfield  {journal} {\bibinfo
  {journal} {Phys. Rev. Lett.}\ }\textbf {\bibinfo {volume} {114}},\ \bibinfo
  {pages} {203001} (\bibinfo {year} {2015})}\BibitemShut {NoStop}%
\bibitem [{\citenamefont {Schmidt}\ and\ \citenamefont
  {Lemeshko}(2016)}]{Schmidt2016}%
  \BibitemOpen
  \bibfield  {author} {\bibinfo {author} {\bibfnamefont {Richard}\ \bibnamefont
  {Schmidt}}\ and\ \bibinfo {author} {\bibfnamefont {Mikhail}\ \bibnamefont
  {Lemeshko}},\ }\bibfield  {title} {\enquote {\bibinfo {title} {Deformation of
  a quantum many-particle system by a rotating impurity},}\ }\href {\doibase
  10.1103/PhysRevX.6.011012} {\bibfield  {journal} {\bibinfo  {journal} {Phys.
  Rev. X}\ }\textbf {\bibinfo {volume} {6}},\ \bibinfo {pages} {011012}
  (\bibinfo {year} {2016})}\BibitemShut {NoStop}%
\bibitem [{\citenamefont {Nielsen}\ and\ \citenamefont
  {Ardila}(2019)}]{nielsen2019critical}%
  \BibitemOpen
  \bibfield  {author} {\bibinfo {author} {\bibfnamefont {K~Knakkergaard}\
  \bibnamefont {Nielsen}}\ and\ \bibinfo {author} {\bibfnamefont {LA~Pe{\~n}a}\
  \bibnamefont {Ardila}},\ }\bibfield  {title} {\enquote {\bibinfo {title}
  {Critical slowdown of non-equilibrium polaron dynamics},}\ }\href@noop {}
  {\bibfield  {journal} {\bibinfo  {journal} {New Journal of Physics}\ }\textbf
  {\bibinfo {volume} {21}},\ \bibinfo {pages} {043014} (\bibinfo {year}
  {2019})}\BibitemShut {NoStop}%
\bibitem [{\citenamefont {Guenther}\ \emph {et~al.}(2021)\citenamefont
  {Guenther}, \citenamefont {Schmidt}, \citenamefont {Bruun}, \citenamefont
  {Gurarie},\ and\ \citenamefont {Massignan}}]{Guenther2021}%
  \BibitemOpen
  \bibfield  {author} {\bibinfo {author} {\bibfnamefont {Nils-Eric}\
  \bibnamefont {Guenther}}, \bibinfo {author} {\bibfnamefont {Richard}\
  \bibnamefont {Schmidt}}, \bibinfo {author} {\bibfnamefont {Georg~M.}\
  \bibnamefont {Bruun}}, \bibinfo {author} {\bibfnamefont {Victor}\
  \bibnamefont {Gurarie}}, \ and\ \bibinfo {author} {\bibfnamefont {Pietro}\
  \bibnamefont {Massignan}},\ }\bibfield  {title} {\enquote {\bibinfo {title}
  {Mobile impurity in a bose-einstein condensate and the orthogonality
  catastrophe},}\ }\href {\doibase 10.1103/PhysRevA.103.013317} {\bibfield
  {journal} {\bibinfo  {journal} {Phys. Rev. A}\ }\textbf {\bibinfo {volume}
  {103}},\ \bibinfo {pages} {013317} (\bibinfo {year} {2021})}\BibitemShut
  {NoStop}%
\bibitem [{big(2022)}]{bighin2022impurity}%
  \BibitemOpen
  \bibfield  {title} {\enquote {\bibinfo {title} {Impurity in a heteronuclear
  two-component bose mixture},}\ }\href@noop {} {\bibfield  {journal} {\bibinfo
   {journal} {Physical Review A}\ }\textbf {\bibinfo {volume} {106}},\ \bibinfo
  {pages} {023301} (\bibinfo {year} {2022})}\BibitemShut {NoStop}%
\bibitem [{\citenamefont {Camacho-Guardian}(2023)}]{camacho2023polaritons}%
  \BibitemOpen
  \bibfield  {author} {\bibinfo {author} {\bibfnamefont {A.}~\bibnamefont
  {Camacho-Guardian}},\ }\bibfield  {title} {\enquote {\bibinfo {title}
  {Polaritons for testing the universality of an impurity in a bose-einstein
  condensate},}\ }\href {\doibase 10.1103/PhysRevA.108.L021303} {\bibfield
  {journal} {\bibinfo  {journal} {Phys. Rev. A}\ }\textbf {\bibinfo {volume}
  {108}},\ \bibinfo {pages} {L021303} (\bibinfo {year} {2023})}\BibitemShut
  {NoStop}%
\bibitem [{\citenamefont {{Mostaan}}\ \emph {et~al.}(2023)\citenamefont
  {{Mostaan}}, \citenamefont {{Goldman}},\ and\ \citenamefont
  {{Grusdt}}}]{Mostaan2023}%
  \BibitemOpen
  \bibfield  {author} {\bibinfo {author} {\bibfnamefont {Nader}\ \bibnamefont
  {{Mostaan}}}, \bibinfo {author} {\bibfnamefont {Nathan}\ \bibnamefont
  {{Goldman}}}, \ and\ \bibinfo {author} {\bibfnamefont {Fabian}\ \bibnamefont
  {{Grusdt}}},\ }\bibfield  {title} {\enquote {\bibinfo {title} {{A unified
  theory of strong coupling Bose polarons: From repulsive polarons to
  non-Gaussian many-body bound states}},}\ }\href {\doibase
  10.48550/arXiv.2305.00835} {\bibfield  {journal} {\bibinfo  {journal} {arXiv
  e-prints}\ ,\ \bibinfo {eid} {arXiv:2305.00835}} (\bibinfo {year} {2023})},\
  \Eprint {http://arxiv.org/abs/2305.00835} {arXiv:2305.00835
  [cond-mat.quant-gas]} \BibitemShut {NoStop}%
\bibitem [{\citenamefont {Paredes}\ \emph {et~al.}(2024)\citenamefont
  {Paredes}, \citenamefont {Bruun},\ and\ \citenamefont
  {Camacho-Guardian}}]{paredes2024perspective}%
  \BibitemOpen
  \bibfield  {author} {\bibinfo {author} {\bibfnamefont {Rosario}\ \bibnamefont
  {Paredes}}, \bibinfo {author} {\bibfnamefont {Georg}\ \bibnamefont {Bruun}},
  \ and\ \bibinfo {author} {\bibfnamefont {Arturo}\ \bibnamefont
  {Camacho-Guardian}},\ }\bibfield  {title} {\enquote {\bibinfo {title}
  {Perspective: Interactions mediated by atoms, photons, electrons, and
  excitons},}\ }\href@noop {} {\bibfield  {journal} {\bibinfo  {journal} {arXiv
  preprint arXiv:2406.13795}\ } (\bibinfo {year} {2024})}\BibitemShut {NoStop}%
\bibitem [{\citenamefont {Bardeen}\ \emph {et~al.}(1967)\citenamefont
  {Bardeen}, \citenamefont {Baym},\ and\ \citenamefont {Pines}}]{Bardeen1967}%
  \BibitemOpen
  \bibfield  {author} {\bibinfo {author} {\bibfnamefont {J.}~\bibnamefont
  {Bardeen}}, \bibinfo {author} {\bibfnamefont {G.}~\bibnamefont {Baym}}, \
  and\ \bibinfo {author} {\bibfnamefont {D.}~\bibnamefont {Pines}},\ }\bibfield
   {title} {\enquote {\bibinfo {title} {Effective interaction of
  ${\mathrm{he}}^{3}$ atoms in dilute solutions of ${\mathrm{he}}^{3}$ in
  ${\mathrm{he}}^{4}$ at low temperatures},}\ }\href {\doibase
  10.1103/PhysRev.156.207} {\bibfield  {journal} {\bibinfo  {journal} {Phys.
  Rev.}\ }\textbf {\bibinfo {volume} {156}},\ \bibinfo {pages} {207--221}
  (\bibinfo {year} {1967})}\BibitemShut {NoStop}%
\bibitem [{\citenamefont {Yu}\ and\ \citenamefont {Pethick}(2012)}]{Yu2012}%
  \BibitemOpen
  \bibfield  {author} {\bibinfo {author} {\bibfnamefont {Zhenhua}\ \bibnamefont
  {Yu}}\ and\ \bibinfo {author} {\bibfnamefont {C.~J.}\ \bibnamefont
  {Pethick}},\ }\bibfield  {title} {\enquote {\bibinfo {title} {Induced
  interactions in dilute atomic gases and liquid helium mixtures},}\ }\href
  {\doibase 10.1103/PhysRevA.85.063616} {\bibfield  {journal} {\bibinfo
  {journal} {Phys. Rev. A}\ }\textbf {\bibinfo {volume} {85}},\ \bibinfo
  {pages} {063616} (\bibinfo {year} {2012})}\BibitemShut {NoStop}%
\bibitem [{\citenamefont {Camacho-Guardian}\ and\ \citenamefont
  {Bruun}(2018)}]{Camacho2018b}%
  \BibitemOpen
  \bibfield  {author} {\bibinfo {author} {\bibfnamefont {A.}~\bibnamefont
  {Camacho-Guardian}}\ and\ \bibinfo {author} {\bibfnamefont {Georg~M.}\
  \bibnamefont {Bruun}},\ }\bibfield  {title} {\enquote {\bibinfo {title}
  {Landau effective interaction between quasiparticles in a bose-einstein
  condensate},}\ }\href {\doibase 10.1103/PhysRevX.8.031042} {\bibfield
  {journal} {\bibinfo  {journal} {Phys. Rev. X}\ }\textbf {\bibinfo {volume}
  {8}},\ \bibinfo {pages} {031042} (\bibinfo {year} {2018})}\BibitemShut
  {NoStop}%
\bibitem [{\citenamefont {Fujii}\ \emph {et~al.}(2022)\citenamefont {Fujii},
  \citenamefont {Hongo},\ and\ \citenamefont {Enss}}]{Fujii2022}%
  \BibitemOpen
  \bibfield  {author} {\bibinfo {author} {\bibfnamefont {Keisuke}\ \bibnamefont
  {Fujii}}, \bibinfo {author} {\bibfnamefont {Masaru}\ \bibnamefont {Hongo}}, \
  and\ \bibinfo {author} {\bibfnamefont {Tilman}\ \bibnamefont {Enss}},\
  }\bibfield  {title} {\enquote {\bibinfo {title} {Universal van der waals
  force between heavy polarons in superfluids},}\ }\href {\doibase
  10.1103/PhysRevLett.129.233401} {\bibfield  {journal} {\bibinfo  {journal}
  {Phys. Rev. Lett.}\ }\textbf {\bibinfo {volume} {129}},\ \bibinfo {pages}
  {233401} (\bibinfo {year} {2022})}\BibitemShut {NoStop}%
\bibitem [{\citenamefont {Drescher}\ \emph {et~al.}(2023)\citenamefont
  {Drescher}, \citenamefont {Salmhofer},\ and\ \citenamefont
  {Enss}}]{Drescher2023}%
  \BibitemOpen
  \bibfield  {author} {\bibinfo {author} {\bibfnamefont {Moritz}\ \bibnamefont
  {Drescher}}, \bibinfo {author} {\bibfnamefont {Manfred}\ \bibnamefont
  {Salmhofer}}, \ and\ \bibinfo {author} {\bibfnamefont {Tilman}\ \bibnamefont
  {Enss}},\ }\bibfield  {title} {\enquote {\bibinfo {title} {Medium-induced
  interaction between impurities in a bose-einstein condensate},}\ }\href
  {\doibase 10.1103/PhysRevA.107.063301} {\bibfield  {journal} {\bibinfo
  {journal} {Phys. Rev. A}\ }\textbf {\bibinfo {volume} {107}},\ \bibinfo
  {pages} {063301} (\bibinfo {year} {2023})}\BibitemShut {NoStop}%
\bibitem [{\citenamefont {{Naidon}}(2018)}]{Naidon2018}%
  \BibitemOpen
  \bibfield  {author} {\bibinfo {author} {\bibfnamefont {Pascal}\ \bibnamefont
  {{Naidon}}},\ }\bibfield  {title} {\enquote {\bibinfo {title} {{Two
  Impurities in a Bose-Einstein Condensate: From Yukawa to Efimov Attracted
  Polarons}},}\ }\href {\doibase 10.7566/JPSJ.87.043002} {\bibfield  {journal}
  {\bibinfo  {journal} {Journal of the Physical Society of Japan}\ }\textbf
  {\bibinfo {volume} {87}},\ \bibinfo {eid} {043002} (\bibinfo {year}
  {2018})},\ \Eprint {http://arxiv.org/abs/1607.04507} {arXiv:1607.04507
  [cond-mat.quant-gas]} \BibitemShut {NoStop}%
\bibitem [{\citenamefont {Camacho-Guardian}\ \emph {et~al.}(2018)\citenamefont
  {Camacho-Guardian}, \citenamefont {Pe\~na Ardila}, \citenamefont {Pohl},\
  and\ \citenamefont {Bruun}}]{Camacho2018a}%
  \BibitemOpen
  \bibfield  {author} {\bibinfo {author} {\bibfnamefont {A.}~\bibnamefont
  {Camacho-Guardian}}, \bibinfo {author} {\bibfnamefont {L.~A.}\ \bibnamefont
  {Pe\~na Ardila}}, \bibinfo {author} {\bibfnamefont {T.}~\bibnamefont {Pohl}},
  \ and\ \bibinfo {author} {\bibfnamefont {G.~M.}\ \bibnamefont {Bruun}},\
  }\bibfield  {title} {\enquote {\bibinfo {title} {Bipolarons in a
  bose-einstein condensate},}\ }\href {\doibase 10.1103/PhysRevLett.121.013401}
  {\bibfield  {journal} {\bibinfo  {journal} {Phys. Rev. Lett.}\ }\textbf
  {\bibinfo {volume} {121}},\ \bibinfo {pages} {013401} (\bibinfo {year}
  {2018})}\BibitemShut {NoStop}%
\bibitem [{\citenamefont {Will}\ \emph {et~al.}(2021)\citenamefont {Will},
  \citenamefont {Astrakharchik},\ and\ \citenamefont
  {Fleischhauer}}]{Will2021}%
  \BibitemOpen
  \bibfield  {author} {\bibinfo {author} {\bibfnamefont {M.}~\bibnamefont
  {Will}}, \bibinfo {author} {\bibfnamefont {G.~E.}\ \bibnamefont
  {Astrakharchik}}, \ and\ \bibinfo {author} {\bibfnamefont {M.}~\bibnamefont
  {Fleischhauer}},\ }\bibfield  {title} {\enquote {\bibinfo {title} {Polaron
  interactions and bipolarons in one-dimensional bose gases in the strong
  coupling regime},}\ }\href {\doibase 10.1103/PhysRevLett.127.103401}
  {\bibfield  {journal} {\bibinfo  {journal} {Phys. Rev. Lett.}\ }\textbf
  {\bibinfo {volume} {127}},\ \bibinfo {pages} {103401} (\bibinfo {year}
  {2021})}\BibitemShut {NoStop}%
\bibitem [{\citenamefont {Petkovi\ifmmode~\acute{c}\else \'{c}\fi{}}\ and\
  \citenamefont {Ristivojevic}(2022)}]{Peto2022}%
  \BibitemOpen
  \bibfield  {author} {\bibinfo {author} {\bibfnamefont {Aleksandra}\
  \bibnamefont {Petkovi\ifmmode~\acute{c}\else \'{c}\fi{}}}\ and\ \bibinfo
  {author} {\bibfnamefont {Zoran}\ \bibnamefont {Ristivojevic}},\ }\bibfield
  {title} {\enquote {\bibinfo {title} {Mediated interaction between polarons in
  a one-dimensional bose gas},}\ }\href {\doibase 10.1103/PhysRevA.105.L021303}
  {\bibfield  {journal} {\bibinfo  {journal} {Phys. Rev. A}\ }\textbf {\bibinfo
  {volume} {105}},\ \bibinfo {pages} {L021303} (\bibinfo {year}
  {2022})}\BibitemShut {NoStop}%
\bibitem [{\citenamefont {Baroni}\ \emph {et~al.}(2024)\citenamefont {Baroni},
  \citenamefont {Huang}, \citenamefont {Fritsche}, \citenamefont {Dobler},
  \citenamefont {Anich}, \citenamefont {Kirilov}, \citenamefont {Grimm},
  \citenamefont {Bastarrachea-Magnani}, \citenamefont {Massignan},\ and\
  \citenamefont {Bruun}}]{baroni2024mediated}%
  \BibitemOpen
  \bibfield  {author} {\bibinfo {author} {\bibfnamefont {Cosetta}\ \bibnamefont
  {Baroni}}, \bibinfo {author} {\bibfnamefont {Bo}~\bibnamefont {Huang}},
  \bibinfo {author} {\bibfnamefont {Isabella}\ \bibnamefont {Fritsche}},
  \bibinfo {author} {\bibfnamefont {Erich}\ \bibnamefont {Dobler}}, \bibinfo
  {author} {\bibfnamefont {Gregor}\ \bibnamefont {Anich}}, \bibinfo {author}
  {\bibfnamefont {Emil}\ \bibnamefont {Kirilov}}, \bibinfo {author}
  {\bibfnamefont {Rudolf}\ \bibnamefont {Grimm}}, \bibinfo {author}
  {\bibfnamefont {Miguel~A}\ \bibnamefont {Bastarrachea-Magnani}}, \bibinfo
  {author} {\bibfnamefont {Pietro}\ \bibnamefont {Massignan}}, \ and\ \bibinfo
  {author} {\bibfnamefont {Georg~M}\ \bibnamefont {Bruun}},\ }\bibfield
  {title} {\enquote {\bibinfo {title} {Mediated interactions between fermi
  polarons and the role of impurity quantum statistics},}\ }\href@noop {}
  {\bibfield  {journal} {\bibinfo  {journal} {Nature Physics}\ }\textbf
  {\bibinfo {volume} {20}},\ \bibinfo {pages} {68--73} (\bibinfo {year}
  {2024})}\BibitemShut {NoStop}%
\bibitem [{\citenamefont {Emin}(2013)}]{emin2013polarons}%
  \BibitemOpen
  \bibfield  {author} {\bibinfo {author} {\bibfnamefont {David}\ \bibnamefont
  {Emin}},\ }\href@noop {} {\emph {\bibinfo {title} {Polarons}}}\ (\bibinfo
  {publisher} {Cambridge University Press},\ \bibinfo {year}
  {2013})\BibitemShut {NoStop}%
\bibitem [{\citenamefont {Fetter}\ and\ \citenamefont
  {Walecka}(1971)}]{Fetter1971}%
  \BibitemOpen
  \bibfield  {author} {\bibinfo {author} {\bibfnamefont {A.L.}\ \bibnamefont
  {Fetter}}\ and\ \bibinfo {author} {\bibfnamefont {J.D.}\ \bibnamefont
  {Walecka}},\ }\href@noop {} {\emph {\bibinfo {title} {Quantum Theory of
  Many-Particle Systems}}},\ Dover Books on Physics Series\ (\bibinfo
  {publisher} {Dover Publications},\ \bibinfo {year} {1971})\BibitemShut
  {NoStop}%
\bibitem [{\citenamefont {{Grusdt}}\ \emph {et~al.}(2024)\citenamefont
  {{Grusdt}}, \citenamefont {{Mostaan}}, \citenamefont {{Demler}},\ and\
  \citenamefont {{Pe{\~n}a Ardila}}}]{Grusdt2024}%
  \BibitemOpen
  \bibfield  {author} {\bibinfo {author} {\bibfnamefont {F.}~\bibnamefont
  {{Grusdt}}}, \bibinfo {author} {\bibfnamefont {N.}~\bibnamefont {{Mostaan}}},
  \bibinfo {author} {\bibfnamefont {E.}~\bibnamefont {{Demler}}}, \ and\
  \bibinfo {author} {\bibfnamefont {Luis~A.}\ \bibnamefont {{Pe{\~n}a
  Ardila}}},\ }\bibfield  {title} {\enquote {\bibinfo {title} {{Impurities and
  polarons in bosonic quantum gases: a review on recent progress}},}\ }\href
  {\doibase 10.48550/arXiv.2410.09413} {\bibfield  {journal} {\bibinfo
  {journal} {arXiv e-prints}\ ,\ \bibinfo {eid} {arXiv:2410.09413}} (\bibinfo
  {year} {2024})},\ \Eprint {http://arxiv.org/abs/2410.09413} {arXiv:2410.09413
  [cond-mat.quant-gas]} \BibitemShut {NoStop}%
\bibitem [{\citenamefont {Sidler}\ \emph {et~al.}(2017)\citenamefont {Sidler},
  \citenamefont {Back}, \citenamefont {Cotlet}, \citenamefont {Srivastava},
  \citenamefont {Fink}, \citenamefont {Kroner}, \citenamefont {Demler},\ and\
  \citenamefont {Imamoglu}}]{sidler2017fermi}%
  \BibitemOpen
  \bibfield  {author} {\bibinfo {author} {\bibfnamefont {Meinrad}\ \bibnamefont
  {Sidler}}, \bibinfo {author} {\bibfnamefont {Patrick}\ \bibnamefont {Back}},
  \bibinfo {author} {\bibfnamefont {Ovidiu}\ \bibnamefont {Cotlet}}, \bibinfo
  {author} {\bibfnamefont {Ajit}\ \bibnamefont {Srivastava}}, \bibinfo {author}
  {\bibfnamefont {Thomas}\ \bibnamefont {Fink}}, \bibinfo {author}
  {\bibfnamefont {Martin}\ \bibnamefont {Kroner}}, \bibinfo {author}
  {\bibfnamefont {Eugene}\ \bibnamefont {Demler}}, \ and\ \bibinfo {author}
  {\bibfnamefont {Atac}\ \bibnamefont {Imamoglu}},\ }\bibfield  {title}
  {\enquote {\bibinfo {title} {Fermi polaron-polaritons in charge-tunable
  atomically thin semiconductors},}\ }\href@noop {} {\bibfield  {journal}
  {\bibinfo  {journal} {Nature Physics}\ }\textbf {\bibinfo {volume} {13}},\
  \bibinfo {pages} {255--261} (\bibinfo {year} {2017})}\BibitemShut {NoStop}%
\bibitem [{\citenamefont {Takemura}\ \emph {et~al.}(2014)\citenamefont
  {Takemura}, \citenamefont {Trebaol}, \citenamefont {Wouters}, \citenamefont
  {Portella-Oberli},\ and\ \citenamefont {Deveaud}}]{takemura2014polaritonic}%
  \BibitemOpen
  \bibfield  {author} {\bibinfo {author} {\bibfnamefont {Naotomo}\ \bibnamefont
  {Takemura}}, \bibinfo {author} {\bibfnamefont {St{\'e}phane}\ \bibnamefont
  {Trebaol}}, \bibinfo {author} {\bibfnamefont {Michiel}\ \bibnamefont
  {Wouters}}, \bibinfo {author} {\bibfnamefont {Marcia~T}\ \bibnamefont
  {Portella-Oberli}}, \ and\ \bibinfo {author} {\bibfnamefont {Beno{\^\i}t}\
  \bibnamefont {Deveaud}},\ }\bibfield  {title} {\enquote {\bibinfo {title}
  {Polaritonic feshbach resonance},}\ }\href@noop {} {\bibfield  {journal}
  {\bibinfo  {journal} {Nature Physics}\ }\textbf {\bibinfo {volume} {10}},\
  \bibinfo {pages} {500--504} (\bibinfo {year} {2014})}\BibitemShut {NoStop}%
\bibitem [{\citenamefont {Levinsen}\ \emph {et~al.}(2019)\citenamefont
  {Levinsen}, \citenamefont {Marchetti}, \citenamefont {Keeling},\ and\
  \citenamefont {Parish}}]{Levinsen2019}%
  \BibitemOpen
  \bibfield  {author} {\bibinfo {author} {\bibfnamefont {Jesper}\ \bibnamefont
  {Levinsen}}, \bibinfo {author} {\bibfnamefont {Francesca~Maria}\ \bibnamefont
  {Marchetti}}, \bibinfo {author} {\bibfnamefont {Jonathan}\ \bibnamefont
  {Keeling}}, \ and\ \bibinfo {author} {\bibfnamefont {Meera~M.}\ \bibnamefont
  {Parish}},\ }\bibfield  {title} {\enquote {\bibinfo {title} {Spectroscopic
  signatures of quantum many-body correlations in polariton microcavities},}\
  }\href {\doibase 10.1103/PhysRevLett.123.266401} {\bibfield  {journal}
  {\bibinfo  {journal} {Phys. Rev. Lett.}\ }\textbf {\bibinfo {volume} {123}},\
  \bibinfo {pages} {266401} (\bibinfo {year} {2019})}\BibitemShut {NoStop}%
\bibitem [{\citenamefont {Bastarrachea-Magnani}\ \emph
  {et~al.}(2019)\citenamefont {Bastarrachea-Magnani}, \citenamefont
  {Camacho-Guardian}, \citenamefont {Wouters},\ and\ \citenamefont
  {Bruun}}]{Bastarrachea2019}%
  \BibitemOpen
  \bibfield  {author} {\bibinfo {author} {\bibfnamefont {M.~A.}\ \bibnamefont
  {Bastarrachea-Magnani}}, \bibinfo {author} {\bibfnamefont {A.}~\bibnamefont
  {Camacho-Guardian}}, \bibinfo {author} {\bibfnamefont {M.}~\bibnamefont
  {Wouters}}, \ and\ \bibinfo {author} {\bibfnamefont {G.~M.}\ \bibnamefont
  {Bruun}},\ }\bibfield  {title} {\enquote {\bibinfo {title} {Strong
  interactions and biexcitons in a polariton mixture},}\ }\href {\doibase
  10.1103/PhysRevB.100.195301} {\bibfield  {journal} {\bibinfo  {journal}
  {Phys. Rev. B}\ }\textbf {\bibinfo {volume} {100}},\ \bibinfo {pages}
  {195301} (\bibinfo {year} {2019})}\BibitemShut {NoStop}%
\bibitem [{\citenamefont {Chin}\ \emph {et~al.}(2010)\citenamefont {Chin},
  \citenamefont {Grimm}, \citenamefont {Julienne},\ and\ \citenamefont
  {Tiesinga}}]{Chin2010}%
  \BibitemOpen
  \bibfield  {author} {\bibinfo {author} {\bibfnamefont {Cheng}\ \bibnamefont
  {Chin}}, \bibinfo {author} {\bibfnamefont {Rudolf}\ \bibnamefont {Grimm}},
  \bibinfo {author} {\bibfnamefont {Paul}\ \bibnamefont {Julienne}}, \ and\
  \bibinfo {author} {\bibfnamefont {Eite}\ \bibnamefont {Tiesinga}},\
  }\bibfield  {title} {\enquote {\bibinfo {title} {Feshbach resonances in
  ultracold gases},}\ }\href {\doibase 10.1103/RevModPhys.82.1225} {\bibfield
  {journal} {\bibinfo  {journal} {Rev. Mod. Phys.}\ }\textbf {\bibinfo {volume}
  {82}},\ \bibinfo {pages} {1225--1286} (\bibinfo {year} {2010})}\BibitemShut
  {NoStop}%
\bibitem [{\citenamefont {Wellein}\ \emph {et~al.}(1996)\citenamefont
  {Wellein}, \citenamefont {R\"oder},\ and\ \citenamefont
  {Fehske}}]{Wellein1996}%
  \BibitemOpen
  \bibfield  {author} {\bibinfo {author} {\bibfnamefont {G.}~\bibnamefont
  {Wellein}}, \bibinfo {author} {\bibfnamefont {H.}~\bibnamefont {R\"oder}}, \
  and\ \bibinfo {author} {\bibfnamefont {H.}~\bibnamefont {Fehske}},\
  }\bibfield  {title} {\enquote {\bibinfo {title} {Polarons and bipolarons in
  strongly interacting electron-phonon systems},}\ }\href {\doibase
  10.1103/PhysRevB.53.9666} {\bibfield  {journal} {\bibinfo  {journal} {Phys.
  Rev. B}\ }\textbf {\bibinfo {volume} {53}},\ \bibinfo {pages} {9666--9675}
  (\bibinfo {year} {1996})}\BibitemShut {NoStop}%
\bibitem [{\citenamefont {Bonc\ifmmode~\breve{}\else \u{}\fi{}a}\ \emph
  {et~al.}(2000)\citenamefont {Bonc\ifmmode~\breve{}\else \u{}\fi{}a},
  \citenamefont {Katras\ifmmode~\breve{}\else \u{}\fi{}nik},\ and\
  \citenamefont {Trugman}}]{Trugman2000}%
  \BibitemOpen
  \bibfield  {author} {\bibinfo {author} {\bibfnamefont {J.}~\bibnamefont
  {Bonc\ifmmode~\breve{}\else \u{}\fi{}a}}, \bibinfo {author} {\bibfnamefont
  {T.}~\bibnamefont {Katras\ifmmode~\breve{}\else \u{}\fi{}nik}}, \ and\
  \bibinfo {author} {\bibfnamefont {S.~A.}\ \bibnamefont {Trugman}},\
  }\bibfield  {title} {\enquote {\bibinfo {title} {Mobile bipolaron},}\ }\href
  {\doibase 10.1103/PhysRevLett.84.3153} {\bibfield  {journal} {\bibinfo
  {journal} {Phys. Rev. Lett.}\ }\textbf {\bibinfo {volume} {84}},\ \bibinfo
  {pages} {3153--3156} (\bibinfo {year} {2000})}\BibitemShut {NoStop}%
\bibitem [{\citenamefont {Bon\ifmmode~\check{c}\else \v{c}\fi{}a}\ and\
  \citenamefont {Trugman}(2001)}]{bonca2001}%
  \BibitemOpen
  \bibfield  {author} {\bibinfo {author} {\bibfnamefont {J.}~\bibnamefont
  {Bon\ifmmode~\check{c}\else \v{c}\fi{}a}}\ and\ \bibinfo {author}
  {\bibfnamefont {S.~A.}\ \bibnamefont {Trugman}},\ }\bibfield  {title}
  {\enquote {\bibinfo {title} {Bipolarons in the extended holstein hubbard
  model},}\ }\href {\doibase 10.1103/PhysRevB.64.094507} {\bibfield  {journal}
  {\bibinfo  {journal} {Phys. Rev. B}\ }\textbf {\bibinfo {volume} {64}},\
  \bibinfo {pages} {094507} (\bibinfo {year} {2001})}\BibitemShut {NoStop}%
\bibitem [{\citenamefont {Macridin}\ \emph {et~al.}(2004)\citenamefont
  {Macridin}, \citenamefont {Sawatzky},\ and\ \citenamefont
  {Jarrell}}]{Macridin2004}%
  \BibitemOpen
  \bibfield  {author} {\bibinfo {author} {\bibfnamefont {A.}~\bibnamefont
  {Macridin}}, \bibinfo {author} {\bibfnamefont {G.~A.}\ \bibnamefont
  {Sawatzky}}, \ and\ \bibinfo {author} {\bibfnamefont {Mark}\ \bibnamefont
  {Jarrell}},\ }\bibfield  {title} {\enquote {\bibinfo {title} {Two-dimensional
  hubbard-holstein bipolaron},}\ }\href {\doibase 10.1103/PhysRevB.69.245111}
  {\bibfield  {journal} {\bibinfo  {journal} {Phys. Rev. B}\ }\textbf {\bibinfo
  {volume} {69}},\ \bibinfo {pages} {245111} (\bibinfo {year}
  {2004})}\BibitemShut {NoStop}%
\bibitem [{\citenamefont {Camacho~Guardian}\ and\ \citenamefont
  {Peña~Ardila}(2024)}]{camacho_guardian_2024_14516566}%
  \BibitemOpen
  \bibfield  {author} {\bibinfo {author} {\bibfnamefont {Arturo}\ \bibnamefont
  {Camacho~Guardian}}\ and\ \bibinfo {author} {\bibfnamefont {Luis}\
  \bibnamefont {Peña~Ardila}},\ }\href {\doibase 10.5281/zenodo.14516566}
  {\enquote {\bibinfo {title} {Dataset: Polaronic dressing of bound states},}\
  } (\bibinfo {year} {2024})\BibitemShut {NoStop}%
\end{thebibliography}%

\end{document}